# Improved numerical plotting of elliptical orbits using radial action coordinates: Has the symmetry of Leibniz's radial theory based on inertia *versus* gravity been ignored?

Ivan R. Kennedy[1], Michael T. Rose[2] and Angus N. Crossan[3]

[1]School of Life and Environmental Sciences, University of Sydney, NSW 2006,

[2]NSW Department of Primary Industries, Wollongbar, NSW, Australia

[3]Quick Test Technologies, c/- Institute of Agriculture, University of Sydney NSW 2006 Australia

**Corresponding author:** Biomedical Building, 1 Central Avenue, Australian Technology Park, Eveleigh, University of Sydney, NSW 2006 Australia, ivan.kennedy@sydney.edu.au

**Abstract**

We show uniquely that two-body gravitational orbits may be plotted conveniently using a rotating radial reference frame rather than the customary Newtonian rectilinear inertial frame. Leibniz, calculus' cofounder and continental contemporary of Newton claimed that the second radial derivative $d^2r/dt^2$ could be found by taking the difference between an inertial force varying inversely with the cubed radius, and gravitational force varying with an inversed squared radius. Leibniz's radial method was severely criticised by Newton and his supporters, preferring a rectilinear inertial frame, claiming Leibniz failed to satisfy Newton's third law of gravity. We show that these two approaches are functionally equivalent if the central body is much larger than that in orbit. Furthermore, we justify Leibniz's least action approach, using the *semilatus rectum* ($l$) of the orbit characteristic of the eccentricity to calculate the Newtonian centripetal acceleration [$(m+M)G/r^2=h^2/lr^2=a_c$], opposed by the inertial acceleration ($h^2/r^3=a_i$), with the net radial acceleration ($a_r=a_i–a_c=d^2r/dt^2$) controlling radius also allowing estimation of angular displacement ($\omega\delta t=\delta\varphi$) and the action for each interval ($mr^2\omega\delta\varphi= mrv\delta s$). Numerically, our Leibniz model can be more accurate since it requires no assumption that linear vectors in the x,y plane adequately simulate a curvilinear trajectory. Our review of Newton's gravitational equation using centre of mass coordinates concludes that the third law is satisfied for symmetry if the gravitational and inertial masses are taken separately as equating force expressions for gravity ($F_g=mMG/r^2=mr_m\omega^2r/l=Mr_M\omega^2r/l$) and radial inertia ($F_i=mr_m\omega^2=Mr_M\omega^2$), each directed to the centre of mass for radial inertia ($mr_m + Mr_M$). This symmetry validates centre of mass radial coordinates for more accurate plotting of all orbits including dual stars, possibly applicable for better understanding of Einstein's relativity. An advantage of radial coordinates is that orbiting couples can also accept perturbations from other nearby bodies acting centripetally ($-h^2/lr^2$), but not inertially.

## 1. Introduction

Since Kepler and Newton established the laws of planetary motion in the 17th century, the mathematical theory enabling the plotting of orbits based on classical gravitation has been prolific [1,2,3]. Can anything of value still be added to this theory?

The main thesis of this paper is that extra clarity regarding the plotting of orbits in classical gravitation can still be obtained using symmetry. This may aid our understanding of the nature of gravitation, still a perplexing problem not adequately explained by Einstein's tensor model of general relativity. While orbits can be plotted mathematically using an exact analytical solution for the two-body problem as ellipses, this is not possible when these orbits are perturbed by the presence of other massive bodies. Depending on the coordinates employed, the orbits of satellites can be plotted by a range of numerical methods. The most common method employs Newton's inertial reference system with Cartesian vectors to plot the orbit, by numerically recalculating the centripetal acceleration for each time interval [3], now in simulations such as MATLAB.

A new numerical method, described here for the first time, relies on the fact that in a rotational reference system the net radial acceleration can be estimated for each interval in the orbit by simply subtracting the centripetal gravitational acceleration inwards, defined partly by the orbital semi-latus rectum, from the inertial acceleration, exerted outwards along the radius. This provides an easily understood means of plotting the orbit as a function of the current radius and the constancy of its angular motion ($h = mr^2\omega$). It is suggested that this novel approach may be extended to more than two-body calculations. The radial action method used here also reveals features of gravity not normally seen.

We anticipate that the ability to consider inertial and gravitational effects as independent may have other practical applications in physical and environmental processes. This is the main reason for our development of this communication, as a basis for future studies in the global environment.

## 2. Methods

### 2.1 *Estimating velocity and acceleration for central forces*

The position of a point *m* in a plane may be determined by Cartesian coordinates ($x, y$) relative to fixed rectangular axes; or, using trigonometry, by polar coordinates ($r$, $\theta$) indicating the distance *r* from the origin *M*, with $\theta$ the angle that the radius vector from the origin to the point makes with the semi-major axis of elliptic motion. This angle in the plane of rotation can also be considered as showing orientation with respect to the distant stars.

A very important physical property of dynamic symmetry ($r^2\omega = h$) has been referred to as the *specific orbital action* [7], being equivalent to the angular momentum per unit mass. Although action has the same physical dimensions as angular momentum ($ML^2T^{-1}$), it corresponds to the actual motion per unit mass through a sector of its trajectory or angle ($\delta\theta$) − effectively a slice of action ($r^2\omega\delta\theta$ or $r\omega.r\delta\theta = r\omega.\delta s$). Angular motion is considered as a ratio between the radius and an arc of the circumference and is therefore dimensionless, but $r^2\omega$ will only have the same value as $r^2\omega\delta\theta$ for one radian of angular motion.

For a central force such as gravitation with constant energy $r^2\omega = h$ is constant, as required by Kepler's observations and his second law of the radius vector sweeping equal areas in equal times. For elliptical orbits, this requires that the infinitesimal changes in $r$ and $\theta$ exactly compensate for each other. That is, the Coriolis acceleration $2dr/dt.d\theta/dt$ is equal and opposite in sign to $rd^2\theta/dt^2$ so that the net normal component of acceleration is zero. This symmetry depends on $r^2\omega$ remaining constant, consistent with a path of least variation in action.

For circular motion where $dr/dt$ and $d^2r/dt^2$ are both zero, just as in the case of a stone rotated on a sling, we have $-r(d\theta/dt)^2$ *or* $-r\omega^2$ as the well-known centripetal acceleration, causing the curved motion towards the centre pivot. For an inertial frame of reference, the centripetal and central accelerations on the stone can be regarded as equal and opposite, so there is no acceleration along the radius $d^2r/dt^2$. In this case the source of the centripetal force giving the circular motion is the tension generated by the musculature in the arm holding the sling.

However, for gravitation in elliptical orbits of varying radius, we will conclude that the term $r\omega^2$ cannot be the actual centripetal acceleration because that should be synonymous with the gravitational radial acceleration equal to $d^2r/dt^2 - r\omega^2$. For a rotating frame of reference, we can introduce an inertial force acting centrifugally outwards, regarded as fictitious by Newton's first law for recto-linear inertial coordinates. It can then be considered that the relative imbalance between the attractive centripetal and the repulsive inertial effects leads to appropriate changes in $d^2r/dt^2$ during the orbit, producing an elliptical orbit. The value of $d^2r/dt^2$ is an observation or a result, not a cause.

## 2.2 *Symmetry in gravitational acceleration*

We draw attention to the characteristic function for a gravitational couple in an elliptical orbit (see Fig. 1 for geometry), the *celerity* ($C = h^2/l$) or quickness of action [7], the product resulting from the combined mass of a couple ($M+m$) and the universal gravitational constant $G$ defined by Newton. From Kepler's second law of equal areas $h/2$ being swept by the radius vector in equal times, we can conclude Equation (1). Here $a$ is the semi-major axis of the ellipse, $b$ the semi-minor axis and $l$ a characteristic parameter known as the *semilatus rectum* (Fig. 1), the normal to the axis $a$ passing through the focus (F or F') near the location of the major body.

$$r^2\omega = b^2\omega_b = a^2\omega_a = l^2\omega_l = h \qquad (1)$$

The period $T$ of the elliptical orbit is equal to the area of the ellipse $\pi ab$ divided by the area swept out per unit time $h/2$, as shown in Equation (2).

$$T = 2\pi ab/h \qquad (2)$$

Substituting for both $a$ and $b$ in this equation from Equation (1), we have

$$T = 2\pi/(\omega_a\omega_b)^{1/2}$$

$$4\pi^2/T^2 = \omega_a\omega_b = \acute{\omega}^2$$

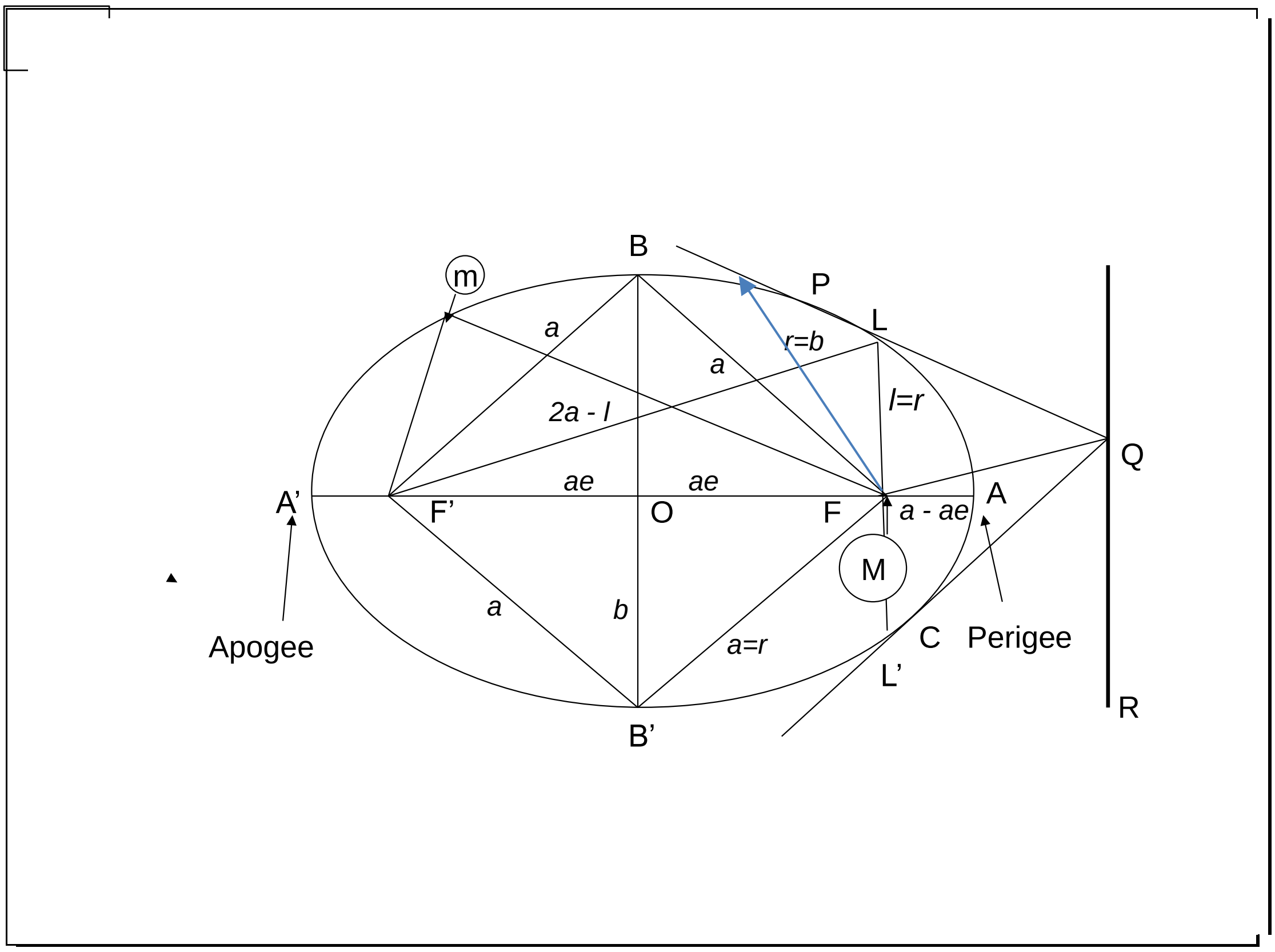

**Figure 1:** Geometrical features of an elliptical orbit (reproduced from [7].

Major semiaxis: AO = gravitational OA' = FB = F'B = FB' = F'B' = $a$

Minor semiaxis: OB = OB' = $b$

Area of ellipse = $\pi ab$ (= $\pi r^2$ when $a = b = r$ in a circle)
OF = OF’ = $a$e (where e is eccentricity)
FL = *semilatus rectum* = $l$ = FL’ = $b^2/a$, $al=b^2$
AF = $a$(1-e) = $l/(1+e)$ = F’A’
A’F = $a$(1+e) = $l/(1-e)$
$1 - e^2 = l/a = b^2/a^2$; $a^2 - b^2 = a^2e^2$
$e = (1 - b^2/a^2)^{0.5}$
FB + BF’= 2$a$ = FP + PF’ = FL + LF’
<FPB’ = <CPB’
<PQF’ = <CQF’

R is a pole on which tangents drawn to the orbit (e.g. QP,QC) are of equal length, the point of their intersection joined to the adjacent focus bisecting the angle between them. If the central body (e.g. the sun) is located at F, then A represents the *perihelion* of the orbit and A’ represents the *aphelion*. As justified in this paper, elliptical orbits are energetically conservative of total energy $E$, with $E = T + V$ equivalent per unit mass equal to $-h^2/2al = -h^2/2b^2 = (dr/dt)^2/2 + h^2/2r^2 - h^2/lr = v^2/2 - h^2/lr$, where $h=r^2\omega$, designated the specific action, is a critical constant for all orbits, related to the conservation of angular momentum.

Here $\acute{\omega}$ is the average angular orbital velocity $2\pi/T$; then Kepler’s third law for the constancy of the product of the cube of the orbital radius with the square of the period is shown in the following equalities for the celerity $C$.

.

$$
\begin{aligned}
4\pi^2a^3/T^2 &= a^3\acute{\omega}^2 = a^3\omega_a\omega_b \\
&= a^3\omega_a^{3/2}\omega_l^{1/2} \\
&= l^3\omega_l^{3/2}\omega_l^{1/2} \\
&= l^3\omega_l^2 \\
&= h^2/l \\
&= C \qquad (3)
\end{aligned}
$$

Given the dimensional form of $G$ ($L^3T^{-2}M^{-1}$), the universal gravitation constant defined by Newton as a function of total mass, we have an exact solution for $C$ in a function defined by the square of the specific action and the semi-latus rectum.

$$G(m + M) = a^3\acute{\omega}^2 = l^3\omega_l^2 = r^4\omega^2/l = h^2/l = C \quad (L^3T^{-2}) \qquad (4)$$

Note that $C$ is not a function solely of the mass of the central body, such as the Sun, but is a function of the total of each system of coupled masses. Where the orbits of satellites of negligible mass compared to the central body are concerned, it is possible to regard $C$ as an approximate constant for these satellites of different masses. However, with modern calculators orbits can be calculated precisely using the actual masses. A computation using an appropriate number of significant digits ─ seven for the planet mercury orbiting the Sun, but at least 28 for a satellite of 100 kg mass in orbit and 57 digits for a proton in solar orbit will be required to appreciate this difference.

In effect, Newton's gravitational constant provides a definition of mass by its effect on celerity $C$. Most significantly, the greater $C$ the greater the straightness of the curvature of the orbit.

**2.3 *Centre of mass coordinates***

We use coordinates based on the centre of masses as ($M$+$m$) lying on $r$ ($r = r_m + r_M$) known as the barycentre; then $mr_m = Mr_M$ and $M = mr_m/r_M$.

Substituting for centre of mass coordinates in Newton's inverse square law of universal gravitation in Equation (5)

$$\begin{aligned} F_g &= -mMG/r^2 \qquad (5) \\ &= -m^2 r_m G/r_M r^2 \\ &= -m^2 r_m h^2/r_M[(m+M)lr^2] \\ &= -m^2 r_m h^2/[(mr_M + mr_m)lr^2] \\ &= -mr_m h^2/lr^3 \\ &= -mr_m \omega^2 r/l \\ &= -Mr_M \omega^2 r/l \qquad (6) \end{aligned}$$

These symmetrical forces ($mr_m\omega^2r/l$ and $Mr_M\omega^2r/l$) can satisfy Newton's third law of action and reaction in the gravitational couple between $m$ and $M$ [5]. Usually, the acceleration involved in the latter force for the solar system $-r_M\omega^2r/l$ is negligible, so close to the Sun's centre is the system's centre of mass, so that $r_m$ is practically equal to $r$ with little or no error in neglecting $r_M$.

However, for complete accuracy in binary systems such as dual stars where $m$ and $M$ are comparable, it becomes imperative to add the two accelerations $-r_M\omega^2r/l$ and $-r_m\omega^2r/l$ for $M$ and $m$ respectively, both directed inwards towards the centre of mass. Together, they will give the total relative acceleration, referred to earlier in Equation (7).

$$r_M\omega^2 r/l + r_m\omega^2 r/l = r^2\omega^2/l \qquad (7)$$

Note that this becomes the same as the gravitational acceleration derived by Newton of $V^2/r$ in circular motion when $r = l$. For elliptical orbits, the gravitational acceleration can also be expressed as $C/r^2$ or $h^2/lr^2$, where $C$ is the celerity $((m+M)G)$, $l$ is the *semilatus rectum* or radial axis normal to the major axis of the ellipse and $h$ is the specific orbital action ($r^2\omega$).

The celerity relationship of $(m + M)G$ equals $h^2/l$ defines any bipolar orbit [6], whether coupling a much larger central body or not. We will show how, for maximum accuracy in plotting, all orbits should be plotted as around the centre of mass of the system.

Another sound reason for using centre of mass coordinates is that this preserves the correct relationship for giving the overall action of the system. Using the centre of mass of one or other always results in an incorrect expression for action. The action is usually given as the integral of the momentum with respect to distance ($\int mv\,ds$). This can be reinterpreted as the integral of $mr^2\omega d\theta$. Given that both members of the couple will change their inertial position during an orbit with respect to the distant stars, the true action for both members for a change of $d\theta$ is $mr_m^2\omega d\theta + Mr_M^2\omega d\theta$ or $mr_m r\omega d\theta = Mr_M r\omega d\theta = I\omega d\theta$. Thus, we have revealed an important symmetry involved in the action of the orbital pairing and how the total action is constant because, $I\omega$ is constant. This cannot be achieved if all of the action is ascribed to the smaller member of the orbital system.

Incidentally, from Equation (6) we can also express the Newtonian gravitational force $F_g$ as follows.

$$
\begin{aligned}
F_g &= -mMG/r^2 \\
&= -mr_m\omega^2 r/l \\
&= -mr_m\omega^2(r_m+r_M)/l \\
&= -(mr^2_m+Mr^2_M)\,\omega^2/l \\
&= -I\omega^2/l
\end{aligned}
$$

Here, $I$ is $(mr^2_m+Mr^2_M)$, the total moment of inertia.

Thus, a re-interpretation of the orbital gravitational force as expressed by Newton is as a variable orbital torque $F_g.l$ proportional to the angular velocity $\omega$ according to the following equation,

$$F_g.l = -I\omega^2 \tag{8}$$

However, this torque is just one-half of that exerted by the couple in the opposite direction, providing a flexible link ensuring that the net torque is zero and there is therefore no change in angular momentum. Note that for a circular orbit where $l = r$, that $F_g$ will be symmetrically equal to $mr_m\omega^2$ and $Mr_M\omega^2$. Under these conditions, the Lagrangian remains constant and there is no variation in action. This also means that the derivative of the total action $I\omega d\theta$ with respect to time is equal to the torque for the orbit, $I\omega^2$, which varies with the angular velocity and the moment of inertia but not with the total angular momentum $I\omega$. This is expected, given that $(m+M)r^2\omega = (m+M)h$.

This symmetry of action is an essential component of gravity.

**2.4 *Cartesian vectors***

The Cartesian differential equations of motion [1] for an attractive force, where $a_c$ is the centripetal or radial acceleration [1], are shown in Equation (9).

$$d^2x/dt^2 = -a_c.cos\theta = -a_c.x/r$$

$$d^2y/dt^2 = -a_c.sin\theta = -a_c.y/r \qquad (9)$$

From Equation (9), the axial components for an acceleration obeying an inverse square law of attraction where $a_c = -h^2/lr^2 = -r^2\omega^2/l$ are given by,

$$d^2x/dt^2 = -Ccos\theta/r^2 = -Cx/r^3 = -h^2x/lr^3 = -x\omega^2r/l = a_x$$

$$d^2y/dt^2 = -Csin\theta/r^2 = -Cy/r^3 = -h^2y/lr^3 = -y\omega^2r/l = a_y$$

$$d^2x/dt^2 + d^2y/dt^2 = -Cx/r^3 - Cy/r^3 = a_x + a_y$$
$$= -(x + y)\omega^2r/l \qquad (10)$$

Note in Figure 2 that the sum of the Cartesian accelerations in the x,y plane is rarely equal to the gravitational acceleration, but usually exceeds it, as ($x + y$) is greater than $r$ (see Fig. 2), except when $x$ or $y$ equals zero. This indicates that resolution of a radial vector into linear Cartesian elements infers that greater forces or impulses must operate than are actually required. This has consequences for plotting orbits increasing with longer time intervals, a point to be examined later in this article. However, when $x = 0$ for the satellite lying on the *latus rectum* where $r = l = y$ (see Fig. 2), it is equal to the gravitational centripetal acceleration. This is also true when $y = 0$ at perigee and apogee in the orbit and $r = x$.

So, to have an equation directly relating the Cartesian accelerations, identified by Newton, and the gravitational acceleration, we must correct Equation (10) by writing down (11).

$$d^2x/dt^2.cos\theta + d^2y/dt^2.sin\theta = -C/r^2(cos^2\theta + sin^2\theta) = -C/r^2 = -r^2\omega^2/l.$$
$$= -G(M+m)/r^2 = -h^2/lr^2 = -h\omega/l$$
$$= d^2r/dt^2 - r\omega^2 \qquad (11)$$

The radius $l$ normal to the major axis of the ellipse where $\theta$ is $\pi/2$ radians is the *semilatus rectum* for the orbit (Fig. 1). The fact that $C$ is equal precisely to $h^2/l$, a unique constant for each stable gravitational couple characteristic of the combined mass $M+m$, is of great significance; the constant $h^2/l$ can be written as the fixed value of $l^3\omega_l^2$ or as $r^3\omega^2.r/l$, a varying relationship to be explored in more detail later.

The $x,y$ plane accelerations of Equations (9, 10) must track as (12) and (13) to equal the centripetal radial acceleration.

$$d^2x/dt^2.cos\theta = d^2x/dt^2.x/r = -G(M+m)x^2/r^4 = -h^2x^2/lr^4 = -x^2\omega^2/l = a_x.x/r \qquad (12)$$

as the vector component of gravitational acceleration along the x-axis, and

$$d^2y/dt^2.sin\theta = d^2y/dt^2.y/r = -G(M+m)y^2/r^4 = -h^2y^2/lr^4 = -y^2\omega^2/l = a_y.y/r \qquad (13)$$

as the vector component of gravitational acceleration along the y-axis.
Obviously, $a_x = -r\omega^2x/l$ and $a_y = -r\omega^2y/l$

Thus, we show in (14) the corrected equivalence for the Cartesian accelerations in Equations (9) and (10).

$$d^2x/dt^2.x/r + d^2y/dt^2.y/r = -G(M+m)/r^2 = -r^2\omega^2/l$$

$$= d^2r/dt^2 - r\omega^2 \qquad (14)$$

The Cartesian accelerations are reduced by the current ratios of $x/r$ and $y/r$ to equal the centripetal acceleration expressed radially.

From Equation (14), we can write energy potentials per unit mass as,

$$x.d^2x/dt^2 + y.d^2y/dt^2 = r.d^2r/dt^2 - r^2\omega^2$$
$$a_x.x + a_y.y = r.a_r - r^2\omega^2 = -r^3\omega^2/l = -C/r$$

When $y$ is equal to both $r$ and $l$ (where the satellite is placed at the *latus rectum* with $x$ equal to zero and the origin as in Figure 2, $d^2y/dt^2$ is equal to $-r\omega^2$, also equal to $-r^2\omega^2/l$. When $x$ is equal to $r$ at apogee or perigee where $y$ is zero, $d^2x/dt^2$ is equal to $-r^2\omega^2/l$ , but not equal to $-r\omega^2$, often called the "centripetal acceleration", since $r$ is not equal to $l$ except twice during an orbit. In fact, the true acceleration of gravitation Newton was seeking is not $-r\omega^2$ but $-r^2\omega^2/l$, except in circular motion when $r = l$ or when the orbit is at the *latus rectum*, the geometric equivalent of circular motion on a conical section.

These Equations (8-13) refer to the sum of the relative accelerations of the satellite towards the central body $-GM/r^2$ with that of the central body towards the satellite $-Gm/r^2$. In most discussions on gravitation with a massive central body $M$, only the former acceleration is considered, neglecting that caused by the satellite $m$ and the component accelerations are then $-GMx/r^3$ and $-GMy/r^3$ [3]. One should be aware that this is merely an approximation. In fact, $-GM/r^2$ is equal to $-r_m{}^2\omega^2/l$ rather than $-r^2\omega^2/l$, where $r_m$ is the radius of $m$ to its centre of mass with $M$ rather than the distance $r$ separating their centres (see Equations (7) and (8)). We can equate $d^2x/dt^2.x/r + d^2y/dt^2.y/r$ to $d^2r_m/dt^2 - r_m\omega^2$, but this is only true where the origin is located at the centre of mass of the gravitational couple $M$ and $m$ rather than the centre of the larger body $M$.

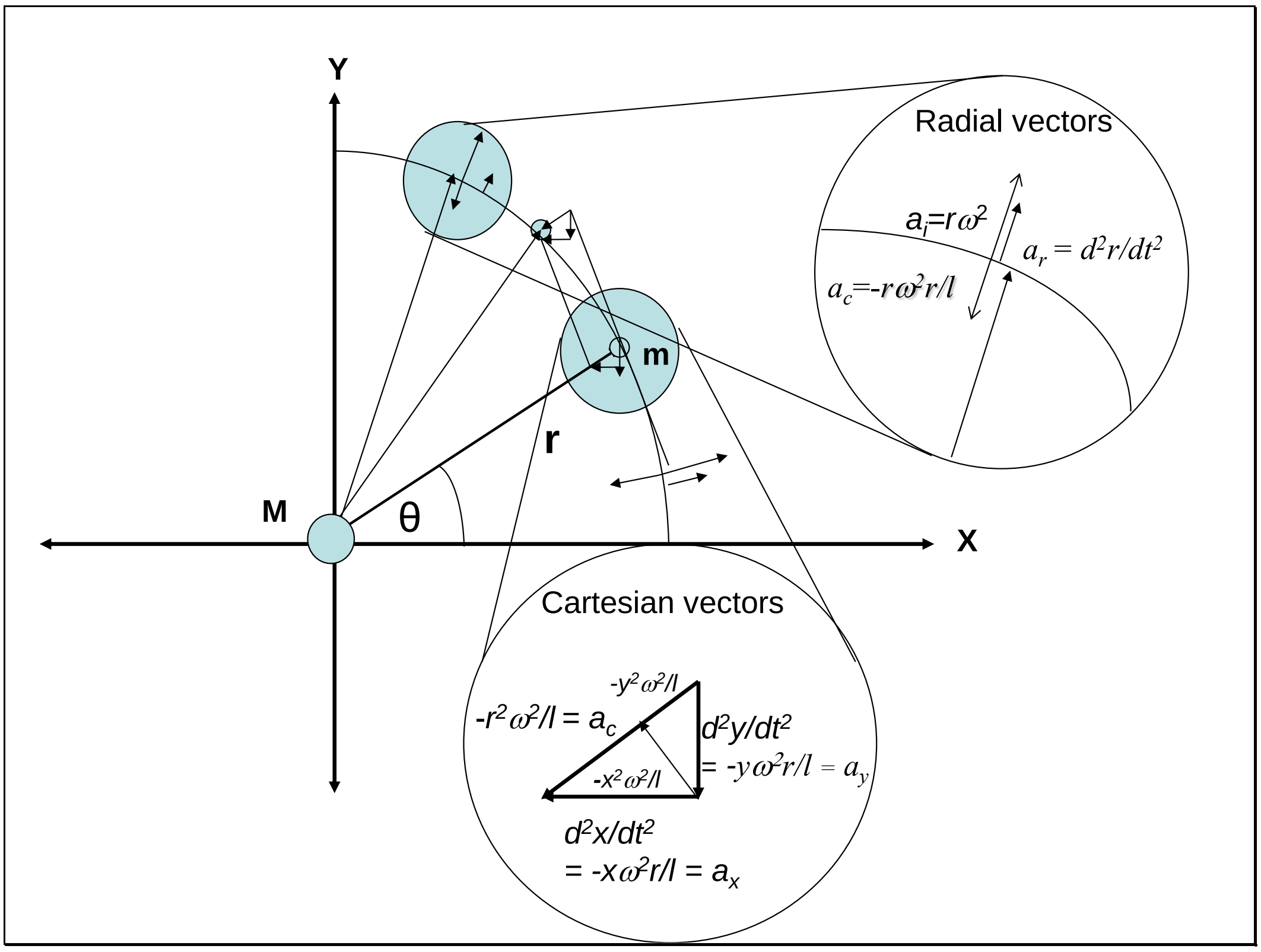

**Figure 2.** The traditional Newtonian plot involves vectors in the $x,y$ plane corresponding to the centripetal acceleration whereas radial vectors operate in a rotating frame, can produce almost the same result. The Cartesian vectors $a_x$ and $a_y$ exceed the radial vector $a_c$ given that $a_x.x/r + a_y.y/r = a_c$ except when x or y equals zero.

Note the equality that the sum of the squares $(dx/dt)^2 + (dy/dt)^2$, given $\sin^2\theta + \cos^2\theta = 1$, is equal to the square of the linear velocity $v$. This is also equal to the radial velocity $(dr/dt)^2$ or $V_r{}^2 + r^2\omega^2$.

$$(dr/dt)^2 + r^2\omega^2 = (dx/dt)^2 + (dy/dt)^2 = v^2$$

Or $V_x^2 + V_y^2 = v^2 = V_r^2 + r^2\omega^2$

## 2.5 *Comparing cartesian versus radial vectors*

Thus, the radial acceleration $d^2r/dt^2$ is actually made of a larger acceleration $d^2r_m/dt^2$ of the smaller satellite *m* towards the centre of mass $r_m$ and a symmetrically opposite acceleration $d^2r_M/dt^2$ of the larger body *M* towards the centre of mass $r_{M.}$ Though acting in opposite directions, these must be added to obtain the total radial acceleration $d^2r/dt^2$ towards the centre of mass. This distinction is usually overlooked, given that it provides a satisfactory approximation when the mass of the central body is orders of magnitude greater than the satellite, as for most bodies in the solar system. But for calculations with dual or binary stars or where the relative masses are less than 100, errors result if this approximation is used.

In this article, the validity of the expression above (21) relating radial $(a_r)$, inertial $(a_i)$ and centripetal accelerations for bodies *m* and *M* bonded by gravitation is examined in detail (Fig. 2). Unlike the case of the stone whirled in a sling, $d^2r/dt^2$ has both positive and negative values during an orbit and the difference is given by the difference between the inertial and the centripetal accelerations. We have shown the validity of the equivalence of variations in radial symmetry using polar coordinates to those using Cartesian coordinates (Fig. 3).

$$d^2r/dt^2 - r.(d\theta/dt)^2 = d^2r/dt^2 - h^2/r^3 = -(M+m)G/r^2 = -h^2/lr^2 = -r^2\omega^2/l \qquad (15)$$

Thus we can write in Equation (16) the relationship involving variation in *dr*/*dt* as one between two symmetries, but out of balance during the orbit except at the semilatus rectum where *l* equals *r*.

$$\begin{aligned} d^2r/dt^2 &= h^2/r^3 - h^2/lr^2 \\ &= r\omega^2 - r\omega^2.r/l = r\omega^2 - r^2\omega^2/l \qquad (16) \\ a_r &= a_i - a_c \end{aligned}$$

Equation (16) provides the basis for calculation of net radial acceleration separate to the use of Cartesian vectors, given that the centripetal acceleration $a_c$ can be calculated from a knowledge of the specific action and the *semilatus rectum.* These two accelerations are different in kind, given that one is inertial derived from the instantaneous circular motion, independent of the influence of the central body, while the other is gravitational and causal, obeying an inverse square relationship. The net radial acceleration ($a_r$) defined here differs from the radial acceleration normally referred to in the theory of gravity of as $d^2r/dt^2 – r\omega^2$; $d^2r/dt^2$ is no more than an observation indicating the rate at which the velocity of the radius vector *dr/dt* varies or reverses in sign at perigee and apogee. We emphasise that the acceleration of the radius $d^2r/dt^2$ is only zero except in a circular orbit whenever the orbit crosses the *latus rectum* when the centripetal and inertial forces are equal. The difference reaches a maximum at perigee when the maxima for both forces is achieved and a minimum at apogee.

The fact that the gravitational $a_c$ and the inertial accelerations $a_i$ are not equal in magnitude does not conflict with Newton's third law, although it was claimed that Newton believed that they should be equal and opposite at all times, according to a recent historical analysis of his *Principia Mathematica* by Aiton [5]. Aiton also indicated that Leibniz recast Newton's fluxions in a form similar to Equation (16). In fact, equality of action and reaction is maintained during orbital motion at all times between both gravitational forces exerted by each member $m$ and $M$ in the gravitational couple on the other; equality is also achieved between the inertial forces exerted by each member of the couple. However, there is no necessity for the gravitational and the inertial forces to be equal to each other unless the orbit is circular, or when $r = l$ as occurs twice in each orbit when the radius vector in normal to the major axis lying on the *latus rectum*. Newton's third law is obeyed, but for the centripetal and the inertial forces between the partners in the couple considered separately and not between these two forces given that they must vary independently during each orbit.

## 3. Results and Discussion

### 3.1 *Symmetrical binding energy potentials*

From Equations (9) and (10) we can conclude that the energy potential per unit mass can be derived from these expressions for the components of acceleration, as follows, given $x^2 + y^2 = r^2$.

$$d^2x/dt^2 + d^2y/dt^2 = -C.x/r^3 - C.y/r^3$$

This relationship for Cartesian accelerations was that used by Feynman [3] to illustrate planetary motions in his lectures at Caltech in 1963. Such an approach allows the motion of all planets simultaneously orbiting the Sun to be calculated using 3-dimensional Cartesian coordinates. Obviously, this process now benefits from the availability of modern computers.

Then to obtain a work function we multiply the coordinate position value by the acceleration, assuming unit mass.

$$\begin{aligned} x.d^2x/dt^2 + y.d^2y/dt^2 &= -C.x^2/r^3 - C.y^2/r^3 = -C(x^2 + y^2)/r^3 \\ &= -C/r = -h^2/lr = -l^3\omega_l^2/r = -r^3\omega^2/l \\ &= -G(M+m)/r \\ &= -GM/r - Gm/r \qquad (17) \end{aligned}$$

It is customary to give the potential energy $V$ for gravitational binding of a satellite as equal to $–GMm/r$ and the energy potential per unit mass as $–GM/r$. However, this neglects the contribution to binding energy of the attraction by the smaller satellite suggested by Equation (23). The energy binding term differs from $-mh^2/lr$ by the ratio $r_m/r$, being equal to $-mr_mr^2\omega^2/l$, where $r_m$ is the length of the radius vector to the centre of mass of $m$ and $M$. Although a very minor error for nearly all the planets, this will no longer be the case when considering the orbits of binary stars, or even the binding potential of small nuclei and electrons. When sufficient precision is available for calculation of orbits, there is no reason why the total energy potential per unit mass bound $-G(M+m)/r$ should not be used, rather than $–GM/r$. This infers that the binding energy is symmetrically derived from each process, not only from binding of the smaller mass.

Within an orbit, the total gravitational energy $E$ is conserved with any change in kinetic energy matched by an opposite change in potential energy. Equation (6) established that the linear kinetic energy per unit mass of $v^2/2$ is equal to $(dr/dt)^2/2 + (r\omega)^2/2$. So with the potential energy from (23) $-G(M+m)/r = -h^2/rl$ we can write Equation (18) for the total energy per unit mass.

$$\begin{aligned} E &= T + V \\ &= -h^2/2al = -h^2/2b^2 = (dr/dt)^2/2 + h^2/2r^2 – h^2/lr \\ &= v^2/2– h^2/lr \qquad (18) \end{aligned}$$

Noted in Figure 1 is the symmetry of how a radius $r_b$ equal to the semi-minor axis $b$ lies between the *semilatus rectum* $l$ and the radius to the minor axis equal to $a$, the semi-major axis. When $r = a$, we have the total potential energy equal to $-h^2/al = -h^2/b^2 = -(r_b\omega_b)^2$, or the negative of the circular velocity squared, consistent with an unchanging total orbital energy $E$ of $-(r_b\omega_b)^2/2$. So, at a radius equal to the semi-minor axis $b$, the linear kinetic energy $v^2/2$ is equal to $h^2/b(1/l – 1/2b)$ and $(dr/dt)^2/2$, the kinetic energy of the change in radius, is equal to $h^2/b(1/l – 1/b)$.

A reinterpretation of the potential energy might be to consider that the $– h^2/lr$ is a static estimate for $V$, given that the circular function for $E$ is constant at $-h^2/2al$ but the circular $T'$ at $r=a$ is $-(r_a\omega_a)^2/2$. The true potential energy could be lessened from $–h^2/lr$ by the increased kinetic energy $(dr/dt)^2/2$, so the result for $V$ is actually $–h^2/lr – (v^2/2 – h^2/2a^2) = h^2/a\,(1/2a - 1/l)$_- $v^2/2$. This would mean that the potential energy around the orbit is diminished from that normally calculated, except at perigee and apogee where $dr/dt$ equals zero. In other words, the potential energy inferred by $-GMm/r$ is for a static satellite $m$ which by necessity it is already diminished by the radial kinetic energy $(dr/dt)^2/2$, unless perigee or apogee is being considered.

In terms of quantised gravitational field energy which is determined by $r$ expressed radially, Equation (19) would be more accurate. But normally this inaccuracy does no harm, as the true zero for potential energy is not used and all arguments about conservation of energy are settled by taking differences.

$$E \quad = v^2/2 - (dr/dt)^2/2 - h^2/lr = h^2/2r^2 - h^2/lr \qquad (19)$$

To place a satellite into orbit it must be given sufficient velocity ($r\omega$) to match the square root of the potential energy in orbit $[h^2/lr]^{0.5}$. This provides the total energy needed for the kinetic energy in orbit $h^2/2r^2$ that could be recovered plus an equal aliquot of field energy thus raising the potential energy twice this value, as required by the virial theorem. Furthermore, if the kinetic energy $h^2/2r^2$ in orbit is then doubled by some form of momentum thrust, the total energy $E$ of gravitational binding will be zero, releasing the orbit entirely. In effect, the negative potential energy $-h^2/lr$ is the loss in potential energy $V$ caused by gravitational bonding, made up of the increase in kinetic energy $h^2/2r^2$ plus the field energy emitted to space on binding, energy previously sustaining the new satellite in some other orbit independently. The zero value assumed is only for the particular process involved. In fact, as we explained in 2001 [7], the trajectories of bound gravitational objects will involve a series of binding processes, each diminishing the original potential energy according to a summation of losses of $\Sigma V$ equal to $[Mc^2 - \Sigma Mv^2]$, comprising the increase in kinetic energy $\Sigma Mv^2/2$ and an equal emission of radiation $\Sigma hv$.

### 3.2 *Plotting gravitational orbits*

Because of the equality of the vectorial expression for the centripetal acceleration and the polar expressions given in Equations (9) and (10 we now have available two related methods of plotting a gravitational orbit. The first of these methods uses a standard vectorial approach of estimating accelerations for the x- and y-axes from a knowledge of $r$, $x$ and $y$ and the equations given above (4, 13-16). A knowledge of the accelerations in the $x$- and $y$-directions then enables calculation of new velocities $v_x$ and $v_y$ after a short time interval and hence a new position $x, y$ and radial separation $r$ can be found using Pythagorus' theorem $x^2 + y^2 = r^2$ and thus the new accelerations determined. The process is reiterated many times to plot a half orbit until $y$ becomes negative (or $\theta$ exceeds $\pi$) and the semi-ellipse simply reflected to complete the orbit.

Alternatively, the orbit can be plotted directly from radial coordinates ($r, \theta$) (Equation 16) again using $r$ to estimate the centripetal acceleration ($a_c$) from Equation (5), then allowing the inertial acceleration ($a_i$) to be found (16), since $a_c.l/r = a_i$. The net radial acceleration ($a_r = d^2r/dt^2$) is easily found by subtraction (see Equation 22). It is then possible to estimate the new $r$, $\omega$ and thus $\theta$, allowing the orbit to be plotted by a process of reiteration until $\theta$ at least reaches $\pi$. This method may be less prone to rounding errors, having fewer calculations, but also makes no assumptions that the elliptical motion is composed of a set of straight lines, depending on the time interval for each reiteration.

We can see from the Equations (4,13-16) that the standard method of plotting an orbit using Newtonian vectors with Cartesian coordinates and the alternative method using accelerations restricted to those along the radius vector subtracting the inertial acceleration with polar coordinates in the infinitesimal limit should yield the same result; they both allow the rate of change in the radius to be estimated. The Cartesian method adheres more closely in principle to Newton's first law, since it appeals only to the central force, inversely proportional to the square of the radius $r^2$ (Equations 2-3). However, the calculus above shows that employing the Cartesian accelerations $a_x$ and $a_y$ to plot the orbits is formally equivalent to the more direct approach of taking the difference between the inertial ($a_i$) and centripetal ($a_c$) accelerations inversely proportional to the cube and the square of the radius respectively and then directly plotting $r$ and $\delta\theta$.

A radial frame of reference thus allows estimation of the variation in $r$ with $\theta$, to establish the orbit, resolving two opposed accelerations. The position of the central body and the satellite can then be considered as stationary with the radius vector increasing in length from perigee to apogee and decreasing from apogee to perigee, similar to a body projected upwards vertically from the earth's surface in its rise and fall, a result of its initial inertia carrying it to the point where its newly acquired potential energy equals its initial kinetic energy, where gravity succeeds in reversing its motion. Furthermore, the whole solar system can be modelled simultaneously using this approach using polar coordinates ($r$, $\theta$, $\varphi$). The centripetal accelerations for all gravitational attractions can be resolved although the Sun will dominate at all times despite temporary perturbations by planets. The sole inertial effect proportional to $r\omega^2$ is dependent on the composite gravitational effects and not on direct interaction with other planets.

It is obvious that the inertial acceleration $h^2/r^3$ or $r^2\omega^2/r = r\omega^2$ is directed along the radius effectively to the stars and necessarily derived from circular motion, normal to the radius. This is not a direct inertial response on the tangent to the linear velocity $v$ demonstrated in Equation (6). Instead, $v$ must be considered as the result of instantaneously circular but stepwise adjustments by quanta regarding kinetic and potential energy, conserving total energy and keeping $h$ constant; changes of $r$ and $\omega$ occur infinitesimally according to the principle of least variation in action. In any conservative system the variation in potential energy is matched by an opposite variation in kinetic energy, keeping their sum constant. This is the requirement for the specific action $h$ to remain constant. This contrasts with the uptake or emission of field energy from the system giving a different orbit, where the variation of total energy and kinetic energy must have the same magnitude but oppsite sign and the change in potential energy is double the change for total energy or kinetic energy, consistent with the virial principle. The total energy is the actual energy causing the field effects, expressed as quanta delivering impulses [7].

### 3.3 *Computer simulations*

In using a microcomputer and numerical methods to plot the orbit — most conveniently beginning from either apogee or perigee on the major axis where $dr/dt$ ($v_r$, cm sec$^{-1}$) and θ (radians) are both zero — it is necessary to consider that the change in $r$ in the first time period $\delta t$ is equal to the average velocity in the period multiplied by $\delta t$, approximately $v_r\delta t/2$ or $a_r\delta t^2/2$. In reiterated loops, the lengthening or shortening of radius vector is then $a_r\delta t^2$, the mean velocity $v_r$ ($a_r\delta t$) multiplied by the subsequent time period. The change in $a_r$ is estimated by recalculating $a_c$ and $a_i$ for each period. Schema for separate computer programs using the radial and vectorial methods to plot elliptical orbits are given in Figures 3 and 4, with all coding in Supplementary Material.

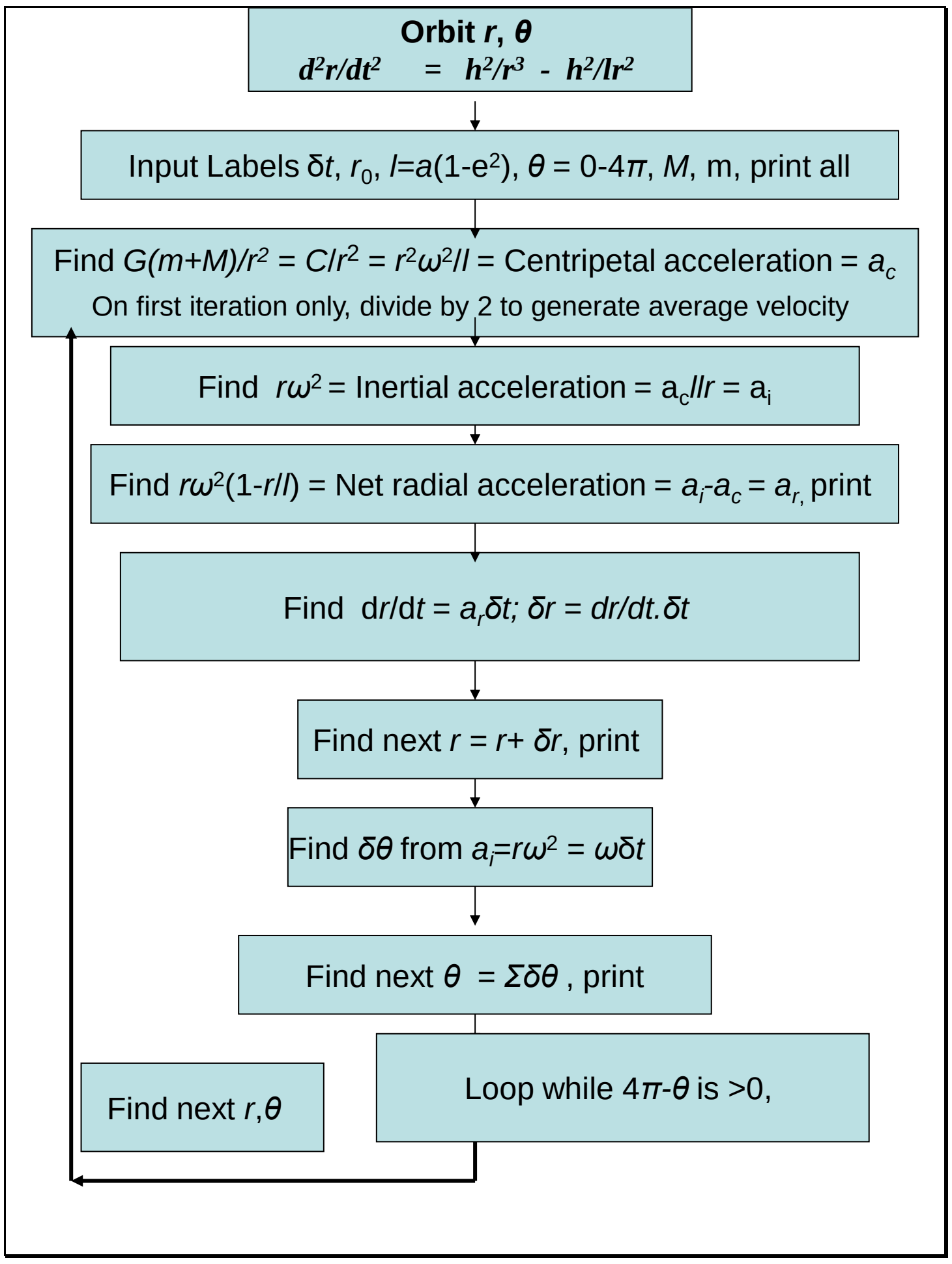

**Figure 3.** Outline of a radial microcomputer program designed to numerically plot a satellite orbit. Note that the first iteration requires an initial multiplicand of 1/2, to obtain the mean velocity in $\delta t_1$, thus avoiding a false radius at the end of period one; all subsequent iterations loop back, omitting this factor. Alternatively, this error in integration could be minimised by shortening $\delta t$.

We present here outlines of two programs (OrbitRadial and OrbitVector) to run numerical calculations, using an interpreter for Windows emulation as TRS32, emulating the Texas Instruments programmable *SR52*; having established the utility of these programs, we also used *R* for Windows, an integrated suite of software facilities for data manipulation, calculation and graphical display available in the public domain. *R* is an implementation of the *S* language which was developed at Bell Laboratories. The code for these programs are given as appendices in supplementary material. Similar programs using centre of mass coordinates ($Mr_{\mathrm{M}} = mr_{\mathrm{m}}$) suitable for plotting orbits for double stars require only minor modification. Another supplementary radial action program we have prepared performs a numerical calculation for the three-body Earth-Moon systems orbiting the Sun. This model can be applied to simulating Newton's Proposition 66, Theorem 26, Case 1, Corr. 4 in his Principia [8] for the Moon's orbit of the Earth in that its motion "is more curved in the quadratures than at the conjunction or the opposition" caused by the Sun's tidal effects when the Moon is on the same radial axis. This observation involving the perturbing gravitational effect of the Sun on the elliptical orbit of the Moon and earth is readily modelled using the radial vector approach.

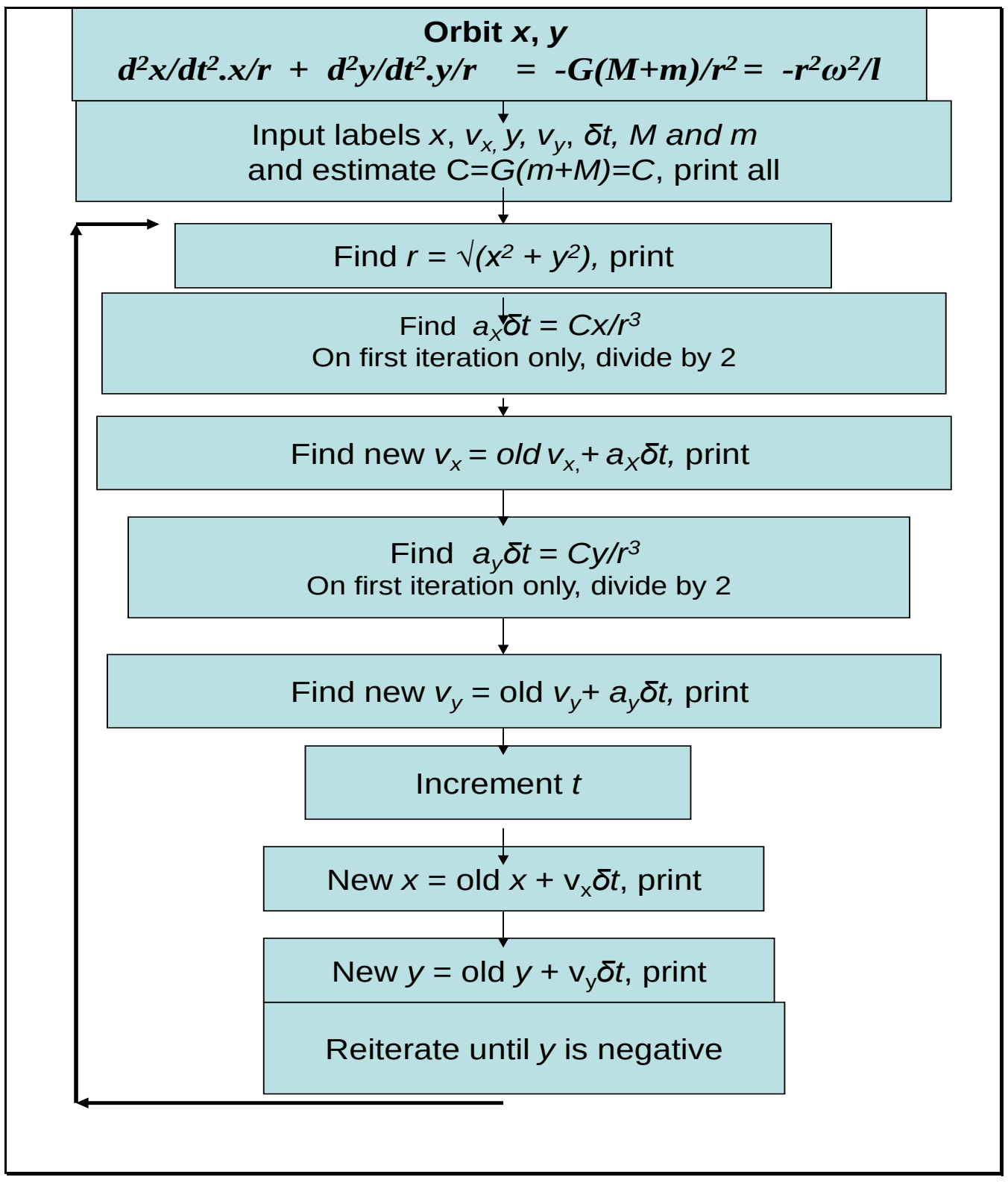

**Figure 4.** Outline of a radial microcomputer program designed to numerically plot a satellite orbit. Note that the first iteration requires an initial multiplicand of 1/2, to obtain the mean velocity in $\delta t_1$, thus avoiding a false radius at the end of period one; all subsequent iterations loop back, omitting this factor. Alternatively, this error in integration could be minimised by shortening $\delta t$.

The Astrocal SR52 program has the advantage for astronomical calculations of much greater accuracy, its 127 significant digits allowing any effect of the mass of the satellite as well as the central body to be included in calculations, despite differences in mass of more than 10 to 100 orders of magnitude between central body and satellite. Also, its operations are highly accessible in practice and the code is more easily developed and understood. But *R* software is readily available to other users since it can be downloaded from the internet free of cost. Full details of these four programs are given in the Appendices to this publication and more information on accessing the software is available on request.

These numerical programs all assume that the change in radius with time is given by the series:

$$\Sigma\delta r = a_{r0}\delta t^2/2 + a_{r1}\delta t^2 + a_{r2}\delta t^2 + a_{rn}\delta t^2 \quad \ldots\ldots \qquad (20)$$

The accelerations ($a_{rn}$) acting are easily calculated from the relationships above and the position of the satellite can then be estimated in either polar or Cartesian coordinates. The change in θ is then easily estimated from the relationship.

$$\delta\theta = (a_i/r_n.)^{0.5}\delta t, \qquad \text{where } a_i = r\omega^2$$

Then $$\Sigma\delta\theta = \Sigma(a_i/r_n.)^{0.5}\delta t = \theta \qquad (21)$$

Given that the specific action or angular momentum per unit mass is a constant as long as total energy is conserved, the relationship of the radial method to least variation of action ($mv\delta s = mr^2\omega\delta\varphi$) is more obvious. For Maupertuis' version of action as the time integral of momentum with respect to differential space, given orbital $mr^2\omega = mh$ is constant with time, then the minimum action per orbit is $2\pi mh$ or $2\pi J$. Any other orbital path proposed must exceed 2π of angular motion. The same result would be obtained by the customary Lagrangian of $\int(T - V)dt$ equals $\int(mh^2/2r^2 + mh^2/rl)dt$ or $J\int\omega(1/2 + r/l)dt$. This action integral implies that most time in the orbit will be spent where $r$ is greater than $l$ with $\omega$ in its minimum values. Rather than least action being least time, the competing forces seek greatest time, where potential energy exceeds kinetic energy, like any pendulum.

Although numerical approaches are an approximation, a very high degree of accuracy is possible if the period $\delta t$ is chosen sufficiently small. There is only slight improvement to accuracy when the number of iterations per orbit is increased beyond 50 or 100. However, the radial and the Cartesian methods do not give precisely the same result in terms of estimating the correct radius at apogee when the same data inputs to the two programs are made at perigee (celerity, radius, masses, etc.), although the agreement is better with smaller time intervals $\delta t$.

The radial method is more accurate as a result of lower error in estimating the change in radius directly from the acceleration directly on the radius vector compared to the Cartesian method requiring separate estimations of changes in $x$ and $y$ first using two finite accelerations at each iteration (Fig. 2).

Some representative outputs from Astrocal for $r$ and θ from data calculated to 79 significant figures for the orbit of the planet Mercury and for hypothetical planet of a range of differing masses are shown in Table 1, allowing a comparison of the radial (Leibnizian) and the Cartesian (Newtonian) models. Also shown are the radii generated for these orbits on the day when a half orbit is completed, close to aphelion. The main limitation to accuracy is that of the input data, but the results for the radial model are highly consistent with expectations, with the *semilatus rectum* reached when θ is $\pi/2$ radians after 18 days and aphelion or maximum radius reached between 87.5 and 88.5 siderial earth days, where θ exceeds $\pi$ radians, just as expected for the planet Mercury.

Initially surprising to the authors, it was found that the two numerical methods do not automatically produce exactly the same outputs. The agreement between the two was excellent as long as the mass of the satellite was negligible compared to that of the Sun, or when the total mass ($m+M$) was kept constant. But when the mass of the hypothetical satellite approached that of the Sun, very large discrepancies are generated between these two reiterative numerical models. So the program recommended by Eisberg and Hyde [4] is not appropriate where $m$ is significant, shown in Table 1 by the discrepancies in the vectorial values of $h$. Cartesian coordinates using the centre of mass as the origin are clearly needed to deal with larger masses or orbiting bodies.

**Table 1:** Orbital properties using radial ($r$θ) and cartesian ($XY$) models for numerical calculation of aphelion ($r_a$), specific action ($h$) and approximate period ($T$), with a radius at perihelion ($r_p$) of 4.6001200x$10^{12}$ cm

| Satellite | Mercury | One-third Sun's mass | Two Suns |
|---|---|---|---|
| $M$ (g) (Sun) | 1.9891x$10^{33}$ | 6.6303x$10^{32}$ | 1.9891x$10^{33}$ |
| $m$ (g) | 3.285x$10^{26}$ | 9.995x$10^{32}$ | 1.9891x$10^{33}$ |
| $r_p$ (cm) x$10^{-12}$ | 4.6001200 | 4.6001200 | 4.6001200 |
| $r_a$ cm, $r$θ x$10^{-12}$ | 6.9636406 | 6.9631531 | 6.9631531 |
| $r_a$ cm, $XY$ x$10^{-12}$ | 7.0100137 | 4.9508116 | 2.1127324 |
| $h = r^2\omega$, x$10^{-19}$ | 2.7132032 | 3.1290185 | 3.8354824 |
| $h = v_y.x - v_x.y$, x$10^{-19}$ | 2.7132032 | 2.7132032 | 2.7132032 |
| $T/2$ days, $r$θ | ca. 44 | ca. 41 | ca. 32 |
| $T/2$ days, $XY$ | ca. 44 | ca. 31 | ca. 14 |

Semi-latus rectum = 5.5405x$10^{12}$ cm

In fact, the large discrepancies in Table 1 are trivial results of the different constraints placed on the orbits using the two methods, radial or polar and rectangular. In the case of the former, the program uses the length of the *semilatus rectum* to constrain the orbit whereas the rectangular method uses the orbital velocity $v_y$ crossing the x-axis at perigee, to constrain the orbit, requiring that the specific action $r^2\omega$ or $h$ should remain constant (see Equation 5). The radial method, more correctly, requires that the specific action should increase with the square root of the extra mass, given that $(M+m)G = h^2/l$ (see Equation 15) and $l$ is held constant for the polar calculations in Table 1, so $h$ must vary instead. In the case of the orbit of two bodies with the Sun's mass, the value of the squared specific action $h^2$ is twice that with one major body, so no other change of inputs is necessary in the radial model.

Orbital plots for mercury and the other cases in Table 1 are shown in Figure 5. An advantage of the radial method is that centre of mass bipolar orbits around can be made around the barycentre with only small modifications of the code, separating $r_m$ and $r_M$. The programmable SR-52 has a special function to interconvert polar and rectangular coordinates using the addressable memory register 00, so Cartesian plots as shown in the figure can readily be executed in the same program. In the plot for Mercury the centre of the Sun is 7.6 km from the barycentre at perigee and 11.5 km at apogee, showing that the Sun is also tracing out an ellipse about 4 km across, based on the common relationship with the semi-latus rectum. Figure 5 also demonstrates the symmetry of orbital motion with mass and the central role of the latus rectum.

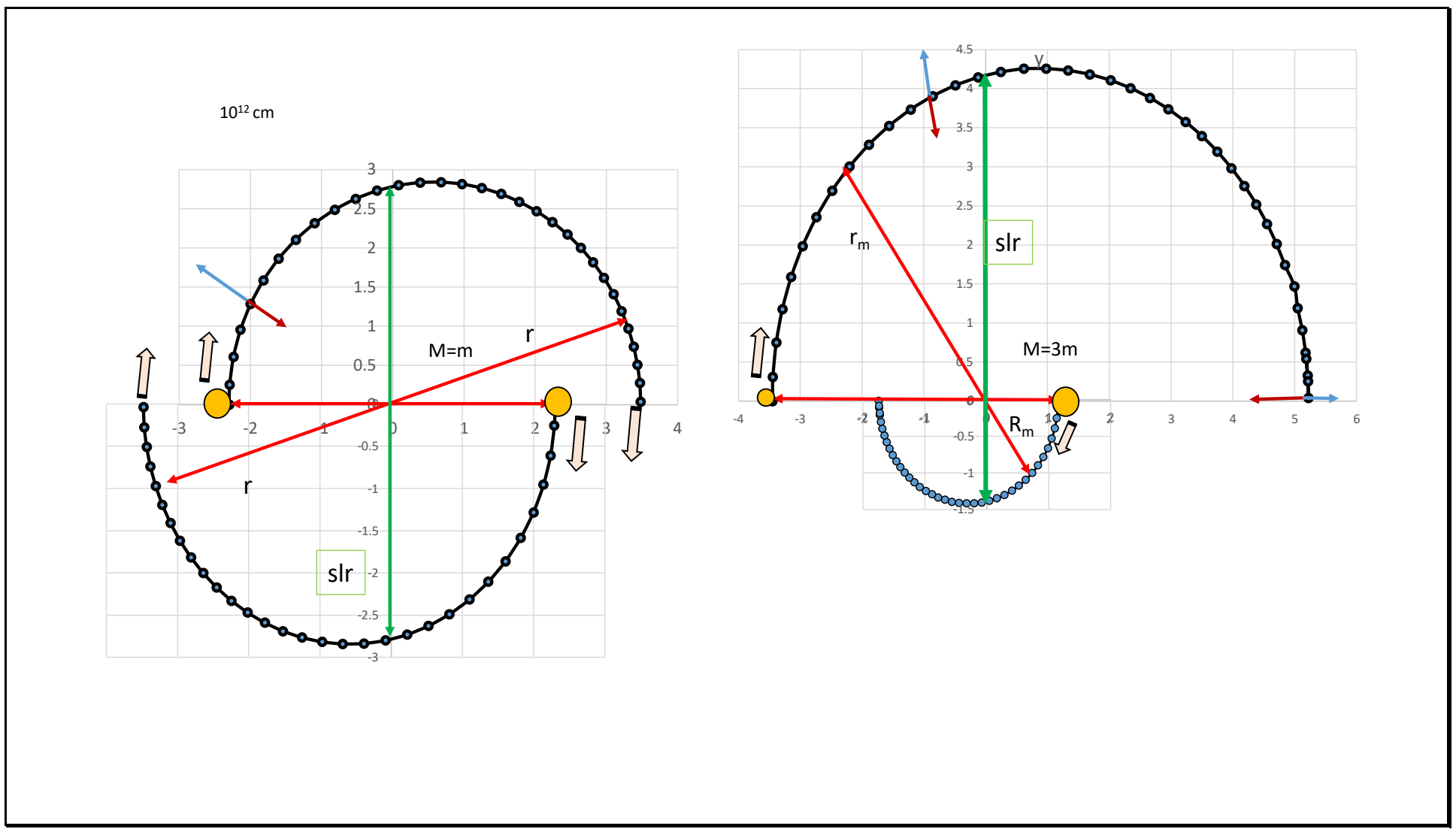

**Figure 5.** Symmetrical radial orbital plots for planet Mercury but substituting masses equal to the Sun and one-third the Sun. For the Sun-Mercury system, only the planetary orbital motion can be shown, given the true barycentre is always within several km of the SUN's centre. Note that the inertial acceleration (blue) is greater than the centripetal or gravitational acceleration at perigee while the reverse is true at apogee. Full data for these plots in Table 1 and for Mercury is given in appendices.

In a simplistic sense, black holes may even be considered as a spherical surface orbiting quanta around a barycentre with a diameter equivalent to *r* with radii $r_o$ equal to the gravitational radius where the celerity *C* or $h^2/l$ is reduced to $r_oc^2$. Newton apparently assumed that gravity was transmitted instantaneously between inertial systems so the concept of a gravitational radius or its derivative identified by Einstein, the Schwarzschild radius was not imaginable.

Given that the radial method produces results that are more accurate when compared with reality as observed with orbits of the Sun's planets (see Table 2), this difference may teach us something about the nature of gravitation so far overlooked. In no case does the polar or radial method lead to an estimate for the radius at aphelion exceeding observations. However, the Cartesian method using rectilinear vectors slightly overestimates both the radius at aphelion and the period of the orbit, although it does give a consistent estimate of the specific action that is exactly in agreement with the data produced by the polar method. This discrepancy in the radius at aphelion is small, being about 1.7% in the case of Mercury and decreasing to 0.48% for Mars and 0.046% for Pluto. But it is significant given the need to obtain maximum accuracy for astronomical purposes. Figure 5 explains why this discrepancy occurs more readily with the Cartesian approach.

**Table 2:** Orbital properties using radial ($r\theta$) and Cartesian (*XY*) models for numerical calculation of aphelion ($r_a$), specific action (*h*) and approximate period, for the planets

| **Satellite** | **Mercury** | **Venus** | **Earth-Moon** | **Mars** | **Pluto** | **Units** |
|---|---|---|---|---|---|---|
| *M* (Sun) | $1.9891\text{x}10^{33}$ | $1.9891\text{x}10^{33}$ | $1.9891\text{x}10^{33}$ | $1.9891\text{x}10^{33}$ | $1.9891\text{x}10^{33}$ | g |
| *M* (Planet) | $3.3022\text{x}10^{26}$ | $4.8685\text{x}10^{27}$ | $5.9737\text{x}10^{27}$ | $6.4185\text{x}10^{26}$ | $1.3050\text{x}10^{25}$ | g |
| $h^2/l$ x$10^{-26}$ | 1.32758126 | 1.32758428 | 1.32758502 | 1.32758146 | 1.32758104 | $\text{cm}^3\text{sec}^{-2}$ |
| $r_p$, x$10^{-12}$ | 4.60012000 | 10.7476259 | 14.70980000 | 20.6669000 | 443.6824613 | cm |
| $r_a$ , $r\theta$ x$10^{-12}$ | 6.9810698 | 10.8941399 | 15.209785041 | 24.9188772 | 737.4297680 | cm |
| $r_a$ *XY* x$10^{-12}$ | 7.0100137 | 10.9111940 | 15.246242910 | 25.03916669 | 740.8509053 | cm |
| $h= r^2\omega$, x$10^{-19}$ | 2.71320322 | 3.79011652 | 4.4558831592 | 5.476451231 | 27.118944372 | $\text{cm}^2\text{sec}^{-1}$ |
| $h= v_y.x–v_x.y$, | 2.71320322 | 3.79011652 | 4.4554558614 | 5.476451231 | 27.118944372 | $\text{cm}^2\text{sec}^{-1}$ |
| $T/2$, $r\theta$ | 44 | 112 | 182 | 340 | Ca. 45000 | days |
| $T/2$, *XY* | 44 | 112 | 182 | 350 | Ca. 46000 | days |
| Actual | 44 | 112 | 182 | 343 | 45260 | |

Accuracy depends largely on the accuracy of the inputs, but the radial method is inherently more accurate.

Calculations made in *R* software have the advantage of giving graphical outputs. Some results for the Mercury-Sun system using the Leibnizian radial approach are shown in Figure 6. Plots indicating the cumulative effect of the radial accelerations in the attractive, centripetal direction ($v_c$) and in the repulsive, inertial direction ($v_i$) are shown. These are obtained from the reiteratively summing the products of the net radial acceleration towards and away from the Sun, being the velocities ($v_r$) with which Mercury is either attracted towards (falling values of $v_r$) or propelled away (increasing values of $v_r$) from the Sun is shown. According to Leibniz's model, these radial vectors are in opposition to each other, with the net radial accelerations and changes in momentum exactly cancelling each other during a complete orbit. The accelerations as indicated by $v_c$ and $v_i$ are of equal magnitude when the satellite is on the *latus rectum*, at right angles to the major axis of the orbit, with the inertial acceleration ($v_i$) being greater at perigee and the centripetal acceleration ($v_c$) in the rotating frame being greater at apogee (see Fig. 6).

The plot in *R* (Fig. 6) shows that the maximum positive and negative values of $v_r$ occur at these positions, as predicted by theory. The corresponding oscillation in the radius of the orbit ($r$) is also shown. The plot for cumulative angular motion is not shown in the figure, but it has exactly the same form as the plot for $v_c$, rather than $v_i$, since the variation in $d\theta/dt$ or angular velocity $\omega$ is inversely proportional to $1/r^2$ for $r^2\omega$ constant during the orbit, just as the variation in $v_c$ is inversely proportional to $1/r^2$. The fact that Newton's centripetal force is directly proportional to the angular velocity $\omega$ no less than $1/r^2$ can hardly be without mechanistic significance.

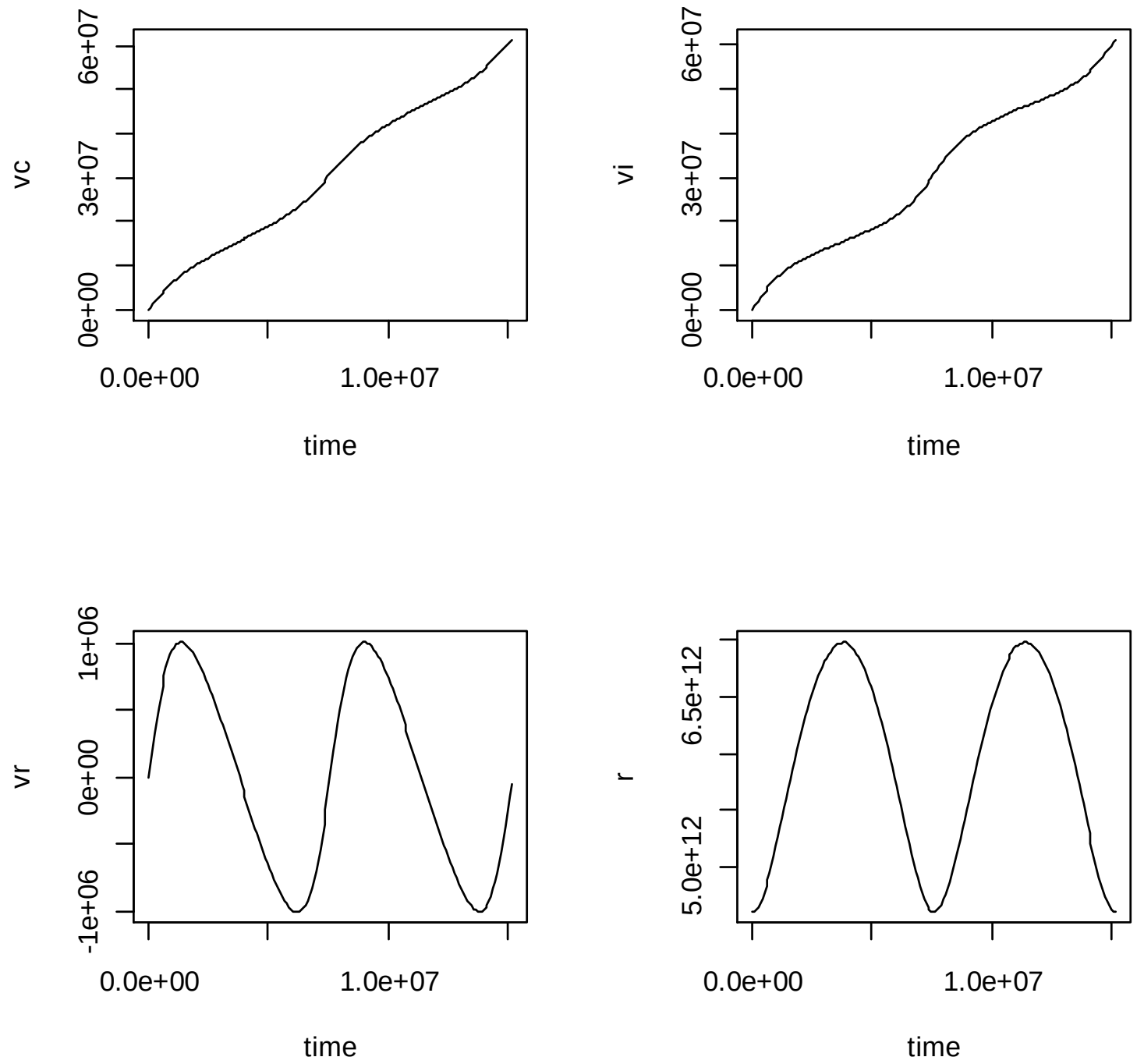

**Figure 6:** Orbital plots from use of Leibnizian radial accelerations to calculate the orbital parameters, centripetal velocity ($v_c$), inertial velocity ($v_i$), radial velocity ($v_r$) radius ($r$), for two orbital periods ($\boldsymbol{\theta}$ = 4π) of the Mercury-Sun system in sec. Note the greater variation in $v_i$ compared to $v_c$, consistent with dependence on inverse radius cubed versus inverse radius squared; $v_r$ is the difference between these two.

The corresponding plots for Mercury-Sun using Cartesian vectors are shown in Figure 7, with the origin ($x$=0, $y$=0) placed at the centre of the Sun and the *latus rectum* being at right angles to the major axis that lies on the x-axis. When $x$ is zero, the magnitude of $dr/dt$ is maximal and it is zero when $y$ is zero, at perigee and apogee.

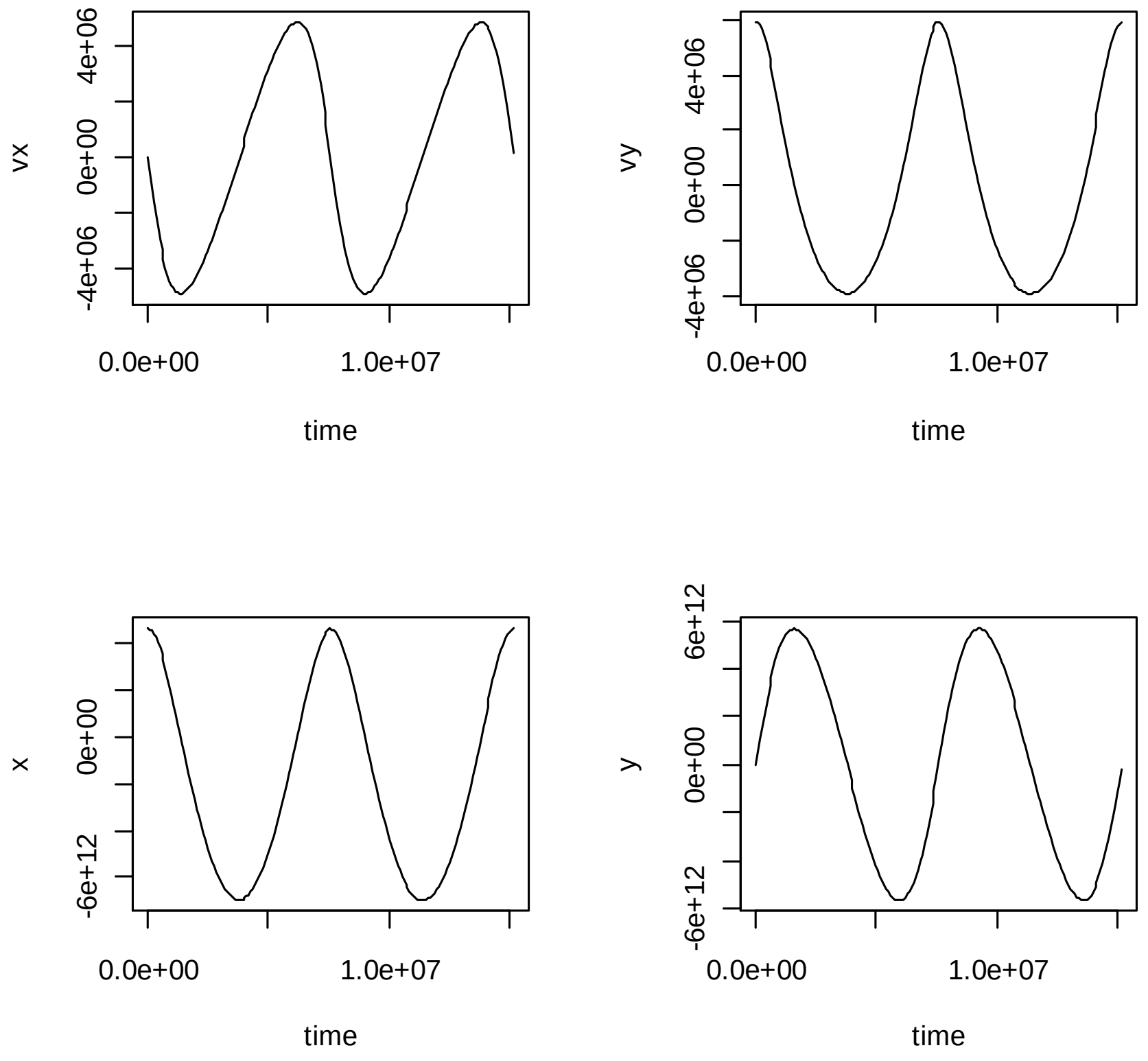

**Figure 7:** Plots from use of Cartesian vectors to calculate the orbital parameters, $x$, $y$ (cm) and $v_x$, $v_y$ (cm/sec) for two orbital periods of the Mercury-Sun system.

Examination of the outputs from *R* by both methods indicate that the results are closely similar, provided the time interval chosen is small (say >50 reiterations per orbit) and within the limits of rounding errors for each method. However, they are not identical.

## 4. Conclusion

We were intrigued to find in October 2010 [4,8], after having written down these relationships in 2005 based on earlier theory [5], that Leibniz had already described the motion of a planet in the 17$^{th}$ century as a contest between the attractive force and the inertial force caused by the rotation, acting to move the body along the rotating radius vector. Leibniz even wrote down one of the first differential equations $d^2r/dt^2 = a/r^3 - b/r^2$, formally equivalent to Equation (16), to describe the resultant acceleration of the radius, where $r$ is the distance from the centre of attraction and $a$ and $b$ are different constants than the semi-major and -minor axes. He drew attention to the significant fact that the inertial force and the centripetal gravitational force were inversely proportional to the cube and the square of the radius respectively.

Leibniz's constants *a* and *b* in his equation would have values of $h^2$ and $h^2/l$ respectively to be considered exactly the same as Equation (16). But it is not clear to the authors if Leibniz considered any significance for the *semilatus rectum*. Newton objected to Leibniz's solution [5, 9], since he claimed the centrifugal force must be equal and opposite to the attraction by his third law of motion, pointing out that Leibniz's equation implied that $d^2r/dt^2$ — logically expressed by Leibniz in his original notation as *ddr/ddt* — was zero. Meli [9] discussed Newton's insistence on his third law as operating between the forces without a clear resolution. However, we have shown that the equality might be better accepted as equations of inertial and gravitational forces between the orbiting bodies around the centre of mass rather than equation of the inertial and centripetal forces. Given Leibniz's role in developing Equation (16) for the second derivative of the radius with time, the fact that he was the first to correct the prevailing Cartesian opinion of the conservation of momentum to the conservation of energy as in the *vis viva* ($mv^2$) is not surprising. Leibniz is also credited with being the first to perceive the role of least action, later taken up by Maupertuis and others.

This paper shows the mathematical equivalence of Newton's idea (discussed with Hooke) that only one gravitational or centripetal force varying with the inverse of the radius squared is needed to produce the orbital motion and Leibniz's idea that it is the result of an ongoing contest between gravitational and inertial effects. We conclude that for the assumption of a rectilinear inertial reference system with a stationary origin (or one moving in a straight line with uniform velocity), Newton is correct; and for a radial frame of reference with the origin located at either the centre of *M* or the centre of mass of *M* and *m*, Leibniz is correct. Newton's reference frame requires Kepler's second law of equal areas and equal times to ensure that the elliptical orbit will be closed whereas Leibniz's rotating frame allows the gravitational acceleration to exceed the inertial acceleration for most of the orbit, causing its eventual closure. Thus, the two points of view differ only in their initial assumptions and can be reconciled. Although similar in principle, these two methods to plot orbits involve subtle differences that require explanation by further analysis. This is despite the facts that both models used exactly the same numerical inputs and the same total accelerations are expected to be exerted from Newton's law of gravity.

An obvious difference is that the radial method makes no assumptions regarding the resolution of linear vectors at each reiterative step required by the Cartesian method to mimic curvilinear motion. However, the inherent inaccuracy using Newton's approach can be much diminished by choosing shorter periods, though at the cost of more computing time. In the radial method, only the acceleration of the radial velocity need be considered, the rotating frame of reference obviating any need to approximate curved trajectories. It should be observed that the net radial acceleration is zero twice during each orbit, when the radius vector is normal to the major axis falling on the *latus rectum*; the radial acceleration reaches its maximum at perihelion with another peak at aphelion, although of lower magnitude and opposite in direction.

The rectangular accelerations given in Equations (9) and (10) are considered very accurate as infinitesimals but average values of these for each period are actually used. But very little additional orbital accuracy is gained by increasing the number of periods beyond the 35-50 used in these simulations for semi-orbits with eccentricity less than 0.25.

For highly elliptical orbits such as that of Halley's comet with an eccentricity of 0.967, it is important when using the numerical methods given in the appendices to shorten the time interval near perihelion and to lengthen it near aphelion, to avoid inaccuracy or too long a plot, no matter whether the polar or the Cartesian method is employed. In the case of Halley's comet, an extremely small proportion of its orbit is occupied by that section where the radius in less than the *semilatus rectum*, so that the centripetal acceleration is nearly always greater than the inertial acceleration, except when the comet lies inside the Earth's orbit close to the Sun. Numerical plots using varying time periods so that $\delta\theta$ is constant for each iterative step would be preferable in such cases, to maximise accuracy and minimise computing time.

Does acceptance of the validity of the Leibnizian approach after all have further implications for gravitational theory? Some physicists consider that rectilinear inertial reference systems are to be preferred as mathematically more correct ─ despite the fact that none can be identified in reality, either stationary or moving in a straight line – even the Sun orbits the galaxy which also orbits its neighbours. Yet they declare that the centrifugal force associated with such curvilinear motion is fictitious. Certainly, it can be agreed that the specific accelerations from gravitation caused by massive bodies and that resulting from the tendency of inertial motion to continue are not in the same class as forces. But inertial accelerations are both real and vary in intensity, as shown by the ability to place satellites in higher orbits using the accelerative thrust of rockets. The inertial tendency to continue in a straight line is clearly a matter of degree and not absolute in effect.

Given that the inertial acceleration directed outwards is calculated continuously as a derivative of the gravitational or centripetal acceleration inwards, there is also the prospect that the influence of many gravitational bodies acting at once can be modelled more effectively in the Leibnizian model used here. Estimated centripetal accelerations can be directed normally and resolved on every significant current radial separation, all diminishing by the inverse square rule. Since the inertial acceleration is exerted normal to the current circular position, it seems to use the distant stars as reference in a Machian sense and nearby objects have no influence on inertia which is self-referencing.

Extending this discussion to relativity is simplified if inertial and gravitational accelerations are considered separately. It is clear that Einstein's general relativity is a function of massive bodies. Yet special relativity concerns inertial effects, though these do interact with gravity. In separate papers (Kennedy and Hodzic, in preparatio) we have proposed that accurate relativistic corrections for GPS clocks can be made with algorithms combining gravitational, with reference to the shortening of space by the gravitational radius $r_o$, including inertial corrections. It may well be that Leibniz's perception that these effects can be considered independently can resonate four centuries later.

**Acknowledgements**

We would like to acknowledge encouragement and criticism from colleagues, particularly James McCaughan (regarding physics) and the late Barrie Fraser (regarding applied mathematics). A supplementary paper in which we focus on least action, written with Barrie Fraser, provides a justification of the radial approach used here by confirming that Leibniz's conclusion is consistent with Newtonian vectorial theory.

**Conflicts of Interest:** The authors declare no conflict of interest.

**Supplementary material**

**Table S1: OrbitRadial**

Program OrbitRadial to calculate $r$, $\theta$ using a Leibnizian radial model by integrating the differential equation $d^2r/dt^2 = h^2/r^3 - h^2/lr^2$ (see Equation 16).

**ASTROCAL CODE**

| Locn. | Code | Instruct. | Comment | Locn. | Code | Intruct. | Comment |
|---|---|---|---|---|---|---|---|
| 000 | 096 | LABEL | | 042 | 096 | LABEL | |
| 001 | 073 | I | | 043 | 087 | W | # Run program |
| 002 | 107 | STO MEM | | 044 | 040 | ( | |
| 003 | 049 | 1 | | 045 | 109 | RCL MEM | |
| 004 | 048 | 0 | # Interval | 046 | 048 | 0 | |
| 005 | 119 | print | δt sec | 047 | 052 | 4 | # M |
| 006 | 002 | HALT | | 048 | 043 | + | |
| 007 | 096 | LABEL | | 049 | 109 | RCL MEM | |
| 008 | 082 | R | | 050 | 048 | 0 | |
| 009 | 107 | STO MEM | | 051 | 053 | 5 | # m |
| 010 | 048 | 0 | | 052 | 041 | ) | |
| 011 | 049 | 1 | # Radius at $t_0$ | 053 | 042 | x | |
| 012 | 119 | print | r cm | 054 | 054 | 6 | # G |
| 013 | 002 | HALT | | 055 | 046 | . | |
| 014 | 096 | LABEL | | 056 | 054 | 6 | |
| 015 | 076 | L | | 057 | 055 | 7 | |
| 016 | 107 | STO MEM | | 058 | 052 | 4 | |
| 017 | 048 | 0 | | 059 | 050 | 2 | |
| 018 | 050 | 2 | # Semilatus rectum | 060 | 056 | 8 | |
| 019 | 119 | print | l cm | 061 | 047 | / | |
| 020 | 002 | HALT | | 062 | 049 | 1 | |
| 021 | 096 | LABEL | | 063 | 101 | EE | # expon. |
| 022 | 084 | T | | 064 | 056 | 8 | |
| 023 | 107 | STO MEM | | 065 | 061 | = | |
| 024 | 048 | 0 | | 066 | 107 | STO MEM | |
| 025 | 051 | 3 | # Angular motion | 067 | 048 | 0 | |
| 026 | 119 | print | $\theta$ radians | 068 | 054 | 6 | # Celerity |
| 027 | 002 | HALT | | 069 | 119 | print | C $cm^3s^{-2}$ |
| 028 | 096 | LABEL | | 070 | 050 | 2 | |
| 029 | 077 | M | | 071 | 114 | 1/x | |
| 030 | 107 | STO MEM | | 072 | 042 | x | |
| 031 | 048 | 0 | | 073 | 109 | RCL MEM | |
| 032 | 052 | 4 | # Mass central body | 074 | 048 | 0 | |
| 033 | 119 | print | M g | 075 | 049 | 1 | # r |
| 034 | 002 | HALT | | 076 | 113 | Square | |
| 035 | 096 | LABEL | | 077 | 114 | 1/x | |
| 036 | 078 | N | | 078 | 042 | x | |
| 037 | 107 | STO MEM | | 079 | 109 | RCL MEM | |
| 038 | 048 | 0 | | 080 | 048 | 0 | |
| 039 | 053 | 5 | # Mass minor body | 081 | 054 | 6 | |
| 040 | 033 | print | m g | 082 | 041 | ) | |
| 041 | 002 | HALT | | 083 | 061 | = | |

**OrbitRadial**

(contnd.)

| Locn. | Code | Instruct. | Comment | Locn. | Code | Intruct. | Comment |
|---|---|---|---|---|---|---|---|
| 084 | 107 | STO MEM | | 125 | 109 | RCL MEM | |
| 085 | 048 | 0 | | 126 | 049 | 1 | |
| 086 | 055 | 7 | # Centripetal | 127 | 050 | 2 | # $v_r$ |
| 087 | 042 | x | acceleration $a_c$ | 128 | 040 | ( | |
| 088 | 109 | RCL MEM | | 129 | 109 | RCL MEM | |
| 089 | 048 | 0 | | 130 | 049 | 1 | |
| 090 | 050 | 2 | # l semilatus rectum | 131 | 049 | 1 | # $\delta v_r$ |
| 091 | 047 | \ | | 132 | 042 | x | |
| 092 | 109 | RCL MEM | | 133 | 109 | RCL MEM | |
| 093 | 048 | 0 | | 134 | 049 | 1 | |
| 094 | 049 | 1 | # r current radius | 135 | 048 | 0 | # $\delta t$ |
| 095 | 041 | ) | | 136 | 041 | ) | |
| 096 | 107 | STO MEM | | 137 | 117 | SUM MEM | |
| 097 | 048 | 0 | # Inertial | 138 | 049 | 1 | |
| 098 | 056 | 8 | acceleration $a_i$ | 139 | 051 | 3 | # $\delta r$ |
| 099 | 109 | RCL MEM | | 140 | 109 | RCL MEM | |
| 100 | 048 | 0 | | 141 | 048 | 0 | |
| 101 | 056 | 8 | # $a_i$ | 142 | 049 | 1 | # r |
| 102 | 045 | - | | 143 | 107 | STO MEM | |
| 103 | 109 | RCL MEM | | 144 | 049 | 1 | |
| 104 | 048 | 0 | | 145 | 052 | 4 | # r old |
| 105 | 055 | 7 | # $a_c$ | 146 | 043 | + | |
| 106 | 061 | = | | 147 | 109 | RCL MEM | |
| 107 | 107 | STO MEM | | 148 | 049 | 1 | |
| 108 | 048 | 0 | | 149 | 051 | 3 | # $\delta r$ |
| 109 | 057 | 9 | # Radial | 150 | 061 | = | |
| 110 | 040 | ( | acceleration $a_r$ | 151 | 107 | STO MEM | |
| 111 | 109 | RCL MEM | | 152 | 048 | 0 | |
| 112 | 048 | 0 | | 153 | 049 | 1 | # r new |
| 113 | 057 | 9 | # $a_r$ | 154 | 119 | print | |
| 114 | 042 | x | | 155 | 109 | RCL MEM | |
| 115 | 109 | RCL MEM | | 156 | 048 | 0 | |
| 116 | 049 | 1 | | 157 | 056 | 8 | # $a_i = r\omega^2$ |
| 117 | 048 | 0 | # $\delta t$ | 158 | 047 | / | |
| 118 | 041 | ) | | 159 | 109 | RCL MEM | |
| 119 | 107 | STO MEM | | 160 | 049 | 1 | |
| 120 | 049 | 1 | | 161 | 052 | 4 | # r old |
| 121 | 049 | 1 | # $\delta v_r$ | 162 | 061 | = | |
| 122 | 117 | SUM MEM | | 163 | 107 | STO MEM | |
| 123 | 049 | 1 | | 164 | 049 | 1 | |
| 124 | 050 | 2 | # $v_r = dr/dt$ | 165 | 053 | 5 | # $\omega^2$ |

NB  The first iteration is a half interval (070-072), to avoid overestimating δr by using the maximum acceleration for the whole interval.

**OrbitRadial**
(contnd.)

| Locn. | Code | Instruct. | Comment |
|---|---|---|---|
| 166 | 092 | Root | |
| 167 | 050 | 2 | |
| 168 | 061 | = | # ω = d$\theta$/dt |
| 169 | 040 | ( | |
| 170 | 107 | STO MEM | |
| 171 | 049 | 1 | |
| 172 | 054 | 6 | # ω |
| 173 | 040 | ( | |
| 174 | 109 | RCL MEM | |
| 175 | 049 | 1 | |
| 176 | 052 | 4 | # r old |
| 177 | 113 | Square | |
| 178 | 042 | x | # $r^2$ old |
| 179 | 109 | RCL MEM | |
| 180 | 049 | 1 | |
| 181 | 054 | 6 | # ω |
| 182 | 041 | ) | # $r^2\omega$ |
| 183 | 119 | print | |
| 184 | 040 | ( | |
| 185 | 109 | RCL MEM | |
| 186 | 049 | 1 | |
| 187 | 054 | 6 | # ω |
| 188 | 042 | x | |
| 189 | 109 | RCL MEM | |
| 190 | 049 | 1 | |
| 191 | 048 | 0 | # δt |
| 192 | 041 | ) | # δ$\theta$ |

| Locn. | Code | Intruct. | Comment |
|---|---|---|---|
| 193 | 043 | + | |
| 194 | 109 | RCL MEM | |
| 195 | 048 | 0 | |
| 196 | 051 | 3 | # $\theta$ old |
| 197 | 061 | = | |
| 198 | 117 | SUM MEM | |
| 199 | 049 | 1 | |
| 200 | 055 | 7 | # $\theta$ new |
| 201 | 109 | RCL MEM | |
| 202 | 049 | 1 | |
| 203 | 055 | 7 | # $\theta$ new |
| 204 | 119 | print | |
| 205 | 040 | ( | |
| 206 | 052 | 4 | |
| 207 | 042 | x | |
| 208 | 112 | pi | # 2 orbits |
| 209 | 045 | - | |
| 210 | 109 | RCL MEM | |
| 211 | 049 | 1 | |
| 212 | 055 | 7 | # $\theta$ new |
| 213 | 041 | ) | |
| 214 | 037 | If pos | |
| 215 | 048 | 0 | |
| 216 | 048 | 0 | |
| 217 | 055 | 7 | |
| 218 | 051 | 3 | # Goto 73 |
| 219 | 002 | HALT | |

The program may be continued to any number of orbits by increasing the factor on location 206.

**Table S2.**
**ASTROCAL CODE OrbitVector**
Program to calculate *x, y* using a Newtonian vectorial model $d^2x/dt^2 + d^2y/dt^2 = -C.x/r^3 - C.y/r^3$ (see Equations (16) and (17).

| Locn. | Code | Instruct. | Comment | Locn. | Code | Intruct. | Comment |
|---|---|---|---|---|---|---|---|
| 000 | 096 | LABEL | | 042 | 096 | LABEL | |
| 001 | 065 | A | | 043 | 078 | N | |
| 002 | 107 | STO MEM | | 044 | 107 | STO MEM | |
| 003 | 048 | 0 | | 045 | 049 | 1 | |
| 004 | 049 | 1 | # x, radius at perigee | 046 | 050 | 2 | # minor mass m |
| 005 | 119 | print | 4.60012x10^12 | 047 | 119 | print | |
| 006 | 002 | HALT | | 048 | 002 | HALT | |
| 007 | 096 | LABEL | | 049 | 096 | LABEL | |
| 008 | 066 | B | | 050 | 080 | P | |
| 009 | 107 | STO MEM | | 051 | 040 | ( | |
| 010 | 048 | 0 | | 052 | 109 | RCL MEM | |
| 011 | 050 | 2 | # $v_x$ at perigee=0 | 053 | 049 | 1 | |
| 012 | 119 | print | | 054 | 049 | 1 | # M |
| 013 | 002 | HALT | | 055 | 043 | + | |
| 014 | 096 | LABEL | | 056 | 109 | RCL MEM | |
| 015 | 067 | C | | 057 | 049 | 1 | |
| 016 | 107 | STO MEM | | 058 | 050 | 2 | # m |
| 017 | 048 | 0 | | 059 | 041 | ) | |
| 018 | 051 | 3 | # y=0 at perigee | 060 | 042 | x | |
| 019 | 119 | print | | 061 | 054 | 6 | # G |
| 020 | 002 | HALT | | 062 | 046 | . | |
| 021 | 096 | LABEL | | 063 | 054 | 6 | |
| 022 | 068 | D | | 064 | 055 | 7 | |
| 023 | 107 | STO MEM | | 065 | 052 | 4 | |
| 024 | 048 | 0 | | 066 | 050 | 2 | |
| 025 | 052 | 4 | # $v_y$=5898657 cm $s^{-1}$ | 067 | 056 | 8 | |
| 026 | 119 | print | | 068 | 047 | / | |
| 027 | 002 | HALT | | 069 | 049 | 1 | |
| 028 | 096 | LABEL | | 070 | 101 | EE | |
| 029 | 069 | E | | 071 | 056 | 8 | |
| 030 | 107 | STO MEM | | 072 | 061 | = | |
| 031 | 048 | 0 | | 073 | 107 | STO MEM | |
| 032 | 053 | 5 | # δt=866164 s | 074 | 049 | 1 | |
| 033 | 119 | print | | 075 | 051 | 3 | # Celerity |
| 034 | 002 | HALT | | 076 | 119 | print | |
| 035 | 096 | LABEL | | 077 | 055 | 2 | |
| 036 | 070 | M | | 078 | 114 | 1/x | # Invert |
| 037 | 107 | STO MEM | | 079 | 042 | x | |
| 038 | 049 | 1 | | 080 | 040 | ( | |
| 039 | 049 | 1 | # Sun's mass M | 081 | 040 | ( | |
| 040 | 119 | print | | 082 | 109 | RCL MEM | |
| 041 | 002 | HALT | | 083 | 048 | 0 | |

NB The first iteration is a half interval (locations 077-079), to avoid overestimating $\delta r$ by using the maximum acceleration for the whole interval.

**OrbitVector**
(contnd.)

| Locn. | Code | Instruct. | Comment |
|---|---|---|---|
| 084 | 041 | 1 | # x |
| 085 | 113 | Square | # $x^2$ |
| 086 | 043 | + | |
| 087 | 109 | RCL MEM | |
| 088 | 048 | 0 | |
| 089 | 051 | 3 | # y |
| 090 | 113 | Square | # $y^2$ |
| 091 | 041 | ( | |
| 092 | 092 | Root | |
| 093 | 050 | 2 | |
| 094 | 041 | ) | |
| 095 | 114 | 1/x | |
| 096 | 091 | Exponent | |
| 097 | 051 | 3 | |
| 098 | 061 | = | # $1/x^3$ |
| 099 | 042 | x | |
| 100 | 109 | RCL MEM | |
| 101 | 049 | 1 | |
| 102 | 051 | 3 | # C |
| 103 | 111 | +/- | # Change sign |
| 104 | 042 | x | |
| 105 | 109 | RCL MEM | |
| 106 | 048 | 0 | |
| 107 | 053 | 5 | # $\delta t$ |
| 108 | 061 | = | # $C\delta t/x^3$ |
| 109 | 107 | STO MEM | |
| 110 | 048 | 0 | |
| 111 | 055 | 7 | # |
| 112 | 042 | x | |
| 113 | 109 | RCL MEM | |
| 114 | 048 | 0 | |
| 115 | 049 | 1 | # x |
| 116 | 061 | = | # $\delta v_x$ |
| 117 | 117 | SUM MEM | |
| 118 | 048 | 0 | |
| 119 | 050 | 2 | # New $v_x$ |
| 120 | 109 | RCL MEM | |
| 121 | 048 | 0 | |
| 122 | 050 | 2 | # $v_x$ |
| 123 | 119 | print | |

| Locn. | Code | Intruct. | Comment |
|---|---|---|---|
| 124 | 109 | RCL MEM | |
| 125 | 048 | 0 | |
| 126 | 055 | 7 | |
| 127 | 042 | x | |
| 128 | 109 | RCL MEM | |
| 129 | 048 | 0 | |
| 130 | 051 | 3 | # y |
| 131 | 061 | = | |
| 132 | 117 | SUM MEM | |
| 133 | 048 | 0 | |
| 134 | 052 | 4 | # New $v_y$ |
| 135 | 109 | RCL MEM | |
| 136 | 048 | 0 | |
| 138 | 052 | 4 | |
| 139 | 109 | RCL MEM | |
| 140 | 048 | 0 | |
| 141 | 053 | 5 | # $\delta t$ |
| 142 | 117 | SUM MEM | |
| 143 | 048 | 0 | |
| 144 | 056 | 8 | # $\delta y$ |
| 145 | 109 | RCL MEM | |
| 146 | 048 | 0 | |
| 147 | 056 | 8 | |
| 148 | 119 | print | |
| 149 | 109 | RCL MEM | |
| 150 | 048 | 0 | |
| 151 | 053 | 5 | # $\delta t$ |
| 152 | 042 | x | |
| 153 | 109 | RCL MEM | |
| 154 | 048 | 0 | |
| 155 | 050 | 2 | # $v_x$ |
| 156 | 061 | = | |
| 157 | 117 | SUM MEM | |
| 158 | 048 | 0 | |
| 159 | 049 | 1 | # New x |
| 160 | 109 | RCL MEM | |
| 161 | 048 | 0 | |
| 162 | 049 | 1 | |
| 163 | 119 | print | # x |

**OrbitVector**
(contnd.)

| Locn. | Code | Instruct. | Comment |
|---|---|---|---|
| 164 | 109 | RCL MEM | |
| 165 | 048 | 0 | |
| 166 | 053 | 5 | # $\delta t$ |
| 167 | 042 | x | |
| 168 | 109 | RCL MEM | |
| 169 | 048 | 0 | |
| 170 | 052 | 4 | # $v_y$ |
| 171 | 061 | = | # $\delta y$ |
| 172 | 117 | SUM MEM | |
| 173 | 048 | 0 | |
| 174 | 051 | 3 | # New y |
| 175 | 109 | RCL MEM | |
| 176 | 048 | 0 | |
| 177 | 051 | 3 | # y |
| 178 | 119 | print | |
| 179 | 040 | ( | |
| 180 | 109 | RCL MEM | |
| 181 | 048 | 0 | |
| 182 | 049 | 1 | # x |
| 183 | 042 | x | |
| 184 | 109 | RCL MEM | |
| 185 | 048 | 0 | |
| 186 | 052 | 4 | # $v_y$ |
| 187 | 041 | ) | |
| 188 | 045 | - | |
| 189 | 040 | ( | |
| 190 | 109 | RCL MEM | |
| 191 | 048 | 0 | |
| 192 | 051 | 3 | # y |
| 193 | 042 | x | |
| 194 | 109 | RCL MEM | |

| Locn. | Code | Intruct. | Comment |
|---|---|---|---|
| 195 | 048 | 0 | |
| 196 | 050 | 2 | # $v_x$ |
| 197 | 041 | ) | |
| 198 | 061 | = | # $h = r^2\omega$ |
| 199 | 109 | RCL MEM | |
| 200 | 048 | 0 | |
| 201 | 049 | 1 | # x |
| 202 | 107 | STO MEM | |
| 203 | 048 | 0 | |
| 204 | 048 | 0 | |
| 205 | 109 | RCL MEM | |
| 206 | 048 | 0 | |
| 207 | 051 | 3 | # y |
| 208 | 064 | Inverse | |
| 209 | 106 | Pol/Rect | |
| 210 | 119 | print | # $\boldsymbol{\theta}$ radians |
| 211 | 109 | RCL MEM | |
| 212 | 048 | 0 | |
| 213 | 048 | 0 | # r cm |
| 214 | 119 | print | # r |
| 215 | 109 | RCL MEM | |
| 216 | 048 | 0 | |
| 217 | 051 | 3 | # y |
| 218 | 037 | If pos | # continues |
| 219 | 048 | 0 | while y is |
| 220 | 048 | 0 | positive |
| 221 | 056 | 8 | |
| 222 | 048 | 0 | |
| 223 | 002 | HALT | |

The program can be run longer than $\pi$ radians by linking the condition in OrbitVectorial location 218 to $\boldsymbol{\theta}$, as in OrbitRadial (locations 205-214).

**Table S3.**
R script for OrbitRadial using differential equation
$dv_r/dt$ = C*l/r^3-C/r^2 where C=$h^2$/l. (see Equation 16).

**R CODE**

**OrbitRadial**

```
require(simecol)

orbitpolar <- new("odeModel",
  main = function(time, init, parms, inputs=NULL) {
    m     <- parms["m"]              # mass of minor body
    M     <- parms["M"]              # mass of major body
    G     <- parms["G"]              # universal gravitational constant
    as    <- parms["as"]             # semi-major axis
    ec    <- parms["ec"]             # eccentricity
    l     <- parms["l"]              # semi-latus rectum
    vc    <- init[1]                 # centripetal velocity
    vi    <- init[2]                 # inertial velocity
    vr     <- init[3]                # net radial velocity
    r     <- init[4]                 # radius
    T     <- init[5]                 # phi in radians to major axis

    l     <- as*(1-ec^2)             # semi-latus rectum
    C     <- (m+M)*G                 # celerity = h^2/l
    h     <- (C*l)^(1/2)             # specific action = r^2dT/dt

          dvi     <- C*l/r^3         # inertial acceleration
          dvc     <- C/r^2           # centripetal acceleration
          dvr  <- C*l/r^3-C/r^2      # net radial acceleration
          dr      <- vi-vc           # radial velocity
          dT      <- (h/r^2)         # angular velocity
          list(c(dvc, dvi, dvr, dr, dT))
 },
```

```
  parms = c(
          m   = 3.3022*10^26,             #  mass of minor body
      M   = 1.9891*10^33,                 #  mass of major body
          G   = 6.67428*10^-8,            #  Gravitational constant
          as  = 5.79091*10^12,            #  Semimajor axis
          ec  = 0.205630,         #  Eccentricity
          l   = 5.5460489*10^12  #  Semilatus rectum
),
  times  = c(from=0, to=86164*176, by=86164),
  init   = c(vc=0, vi=0, vr=0, r=4.60012*10^12, T=0),
  solver = "lsoda"
)
orbpol <- sim(orbitpolar)       # initial simulation
plot(orbpol)
print(orbpol, all=TRUE)         # returns all values in the all "slots"
```

**Table S4**

R script for OrbitVector integrating differential equations

$dv_x/dt = -C*x/((x\hat{}2+y\hat{}2)\hat{}(3/2))$ and $dv_y/dt = -C*y/((x\hat{}2+y\hat{}2)\hat{}(3/2))$,

or $dv_x/dt = -Cx/r^3$ and $dv_y/dt = -Cy/r^3$ (see Equations (16) and (17)).

**R**

**OrbitVector**

```
require(simecol)
orbitvector <- new("odeModel",

  main = function(time, init, parms, inputs=NULL) {
    m     <- parms["m"]                    # mass of minor body
    M     <- parms["M"]                    # mass of major body
    G     <- parms["G"]                    # universal gravitational constant
    C     <- parms["C"]                    # celerity = (m+M)G
```

```
  x              <- parms["x"]           # x-vector distance
  y              <- parms["y"]           # y-vector distance
  vx   <- parms["vx"]                    # x-vector velocity
  vy     <- parms["vy"]                  # y-vector velocity

  vx   <- init[1]
  vy   <- init[2]
  x    <- init[3]
  y    <- init[4]

  C    <- (m+M)*G

   dvx<- -C*x/((x^2+y^2)^(3/2))
   dvy<- -C*y/((x^2+y^2)^(3/2))
   dx<-  vx
   dy<-  vy

   list(c(dvx, dvy, dx, dy))
  },

 parms = c(
        m   = 3.3022*10^26,              #  mass of minor body
        M   = 1.9891*10^33,              #  mass of major body
        G   = 6.67428*10^-8,             #  Gravitational constant
        l   = 5.5460489*10^12            #  Semilatus rectum
),
 times = c(from=0, to=86164*176, by=86164),
 init  = c(vx=0, vy=5898657, x= 4.60012*10^12, y=0),
 solver = "lsoda"
 )
orbvec <- sim(orbitvector)               # initial simulation
plot(orbvec)
print(orbvec, all=TRUE)                  # returns all values in the all "slots"
```